\newcommand{\Msun}{\mathrm{M}_{\odot}}

\documentclass[twocolumn]{aastex63}

\received{}
\revised{}
\accepted{}
\submitjournal{ApJ}

\shorttitle{}
\shortauthors{Deme et al.}
\graphicspath{{./}{figures/}}

\usepackage{amsmath}

\begin{document}

\title{Intermediate mass black holes' effect on compact object binaries}

\correspondingauthor{Barnab{\'a}s Deme}
\email{deme.barnabas@gmail.com}

\author[0000-0003-4016-9778]{Barnab{\'a}s Deme}
\affiliation{Institute of Physics, E\"otv\"os University, P\'azm\'any P. s. 1/A, Budapest, 1117, Hungary}

\author[0000-0003-3518-5183]{Yohai Meiron}
\affiliation{Institute of Physics, E\"otv\"os University, P\'azm\'any P. s. 1/A, Budapest, 1117, Hungary}
\affiliation{Department of Astronomy and Astrophysics, University of Toronto, 50 Saint George St, Toronto, ON M5S 3H4, Canada}

\author[0000-0002-4865-7517]{Bence Kocsis}
\affiliation{Institute of Physics, E\"otv\"os University, P\'azm\'any P. s. 1/A, Budapest, 1117, Hungary}

\begin{abstract}

Although their existence is not yet confirmed observationally, intermediate mass black holes (IMBHs) may play a key role in the dynamics of galactic nuclei. In this paper, we neglect the effect the nuclear star cluster itself and investigate only how a small reservoir of IMBHs influences the secular dynamics of stellar-mass black hole binaries, using $N$-body simulations. We show that our simplifications are valid and that the IMBHs significantly enhance binary evaporation by pushing the binaries into the Hill-unstable region of parameter space, where they are separated by the SMBH's tidal field. For binaries in the S-cluster region of the Milky Way, IMBHs drive the binaries to merge in up to $1$--$6\%$ of cases, assuming five IMBHs within 5 pc of mass $10^4~\Msun$ each. Observations of binaries in the Galactic center may strongly constrain the population of IMBHs therein. 

\end{abstract}


\section{Introduction}

Galactic nuclei are dense stellar environments in the central parsec of galaxies, hosting several important astrophysical phenomena. Supermassive black holes (SMBH) reside in their centers, which are surrounded by a complex structure of gas, stars and stellar mass black holes \citep{genzel2010,neumayer2020}. Many of those stellar mass black holes may be members of binaries \citep{Hailey2018}, whose secular evolution is affected by several factors, including Lidov-Kozai oscillations, precession induced by general relativistic effects or by the mass of the stars enclosed by the orbit around the massive black hole, stellar encounters, two-body and resonant relaxation, non-sphericity of the nuclear stellar cluster, etc. \citep{antonini2012,pfuhl2014,alexander2017,petrovich2017}. These effects may collectively drive the binaries to merge or disrupt \citep{stephan2017,hoang2018,hamers2018}. By the latter we mean that the binary breaks apart and its members continue to orbit independently around the SMBH thereafter, and not that the individual stars suffer tidal disruption. 

Lidov-Kozai oscillations operate in hierarchical three body configurations consisting of a tight `inner' binary (e.g. a stellar mass black hole binary with total mass between $\sim 10$ and $\sim100~\Msun$) whose barycenter revolves around a third body (i.e. in this study an SMBH, with mass between $\sim 10^5$ and $\sim10^{10}~\Msun$) on a much wider orbit. In this case, the eccentricities and the mutual inclination between the inner and outer orbital planes exhibit quasi-periodic variations at fixed semi-major axes (see \citealt{naoz2016} for a review and \citealt{hamilton1,hamilton2} for a more general approach of the phenomenon). The Lidov-Kozai mechanism is especially efficient if the initial mutual inclinations is close to $90^{\circ}$. In the case of an eccentric outer orbit (i.e. the eccentric Lidov-Kozai mechanism), the inner eccentricity can be excited up to extreme values very close to unity for a wide range of initial inclination, as the (chaotic) dynamics is dominated by the octupole-order perturbation \citep{lithwick}. If the inner binary consists of BHs and/or neutron stars (NSs), the very close approach at periapsis can result in either a gamma ray burst detectable with electromagnetic observatories or a powerful emission of gravitational waves that leads to merger, potentially detectable with existing and future detectors such as LIGO\footnote{\url{https://www.ligo.org/}}, VIRGO\footnote{\url{http://www.virgo-gw.eu/} }, KAGRA\footnote{\url{https://gwcenter.icrr.u-tokyo.ac.jp/en/}}, and LISA\footnote{\url{https://lisa.nasa.gov/}} \citep{antonini2012,stephan2017,hoang2018,hamers2018}. The mergers rates can be up to $10^{-1} \mbox{ Gpc}^{-3}\mbox{ yr}^{-1}$ \citep{fragione_b}, or even 5-8 times higher if we consider triples of compact objects instead of binaries \citep{fragione_a}.

Galactic nuclei may also host intermediate mass black holes (IMBH) with mass between $\sim10^3$ and $10^4~\Msun$ beside the  supermassive (SMBH) and the stellar mass black holes. Their existence is not yet observationally confirmed, but there are several candidates (see \citealt{mezcua2017} for a review). Possible theoretical scenarios for their origin include formation from very massive Population III stars \citep{madau2001}, runaway mergers in dense clusters \citep{pzwart2006}, dynamical interactions of binaries containing a stellar mass black hole \citep{giersz2015} or formation in accretion disks around SMBHs \citep{goodman2004,mckernan2012,McKernan2014}. If IMBHs exist, they may have important effects on the dynamics of the galactic nucleus \citep{yu2003,MBattisti2014,arcaseddaaskar,wang}. \citet{girma2018} predicted astrometric biases in the position and proper motion of the central massive black hole and the nuclear star cluster induced by IMBHs, which might be detectable by the next generation of telescopes (see also \citealt{gualandris2009} and \citealt{gualandris2010}). 
If these IMBHs formed in globular clusters that sank into the Galactic center to within 10-100 mpc of the SMBH via dynamical friction, then the observed distribution of the S-stars may be explained by this mechanism (\citealt{merritt2009,arcasedda2018}, however see \citealt{MBattisti2014}). 
Such scenarios result in a few IMBHs in the central parsecs \citep{pzwart2006}, however, their mass distribution and overall number is still highly uncertain. Their presence in the Galactic Center may be observationally tested in the future with pulsar timing \citep{kocsis2012}.

In this paper, we investigate the impact of IMBHs on the dynamics of a compact object binary (COB, including either black holes or neutron stars) around a central supermassive black hole (SMBH) with direct N-body simulations. In particular we examine how the IMBHs affect the Lidov-Kozai oscillations of COBs and quantify the fraction of binaries that are destroyed, i.e. either disrupted or merged together. We do not take into account the dynamical friction of IMBHs on the cluster stars, which makes our results somewhat heuristic but does not invalidate them because it is not efficient at the radii of our interest. We will show that IMBHs significantly decrease the survival probability of COBs in galactic nuclei, and hence their presence puts a strong constraint on whether or not a nucleus contains IMBHs (for a similar investigation see \cite{leigh}, where the central massive object was considered to be an IMBH). 

COB mergers are of special interest since the beginning of gravitational wave astronomy \citep{o2}. However, the evolution and mergers of COBs in galactic nuclei have been previously examined by neglecting the effects of IMBHs  \citep{Hopman2009,pfuhl2014,antonini2012,stephan2016,Petrovich_Antonini2017,hoang2018,hamers2018,hamers2019,Fragione_Ginsburg_Loeb2019,Fragione_Antonini2019,Trani_Fujii_Spera2019}. 

The paper is organized as follows. In Section \ref{sec:numericalsetup} we describe the adopted model of galactic centers, COBs, and IMBHs. In Section \ref{sec:techniques} we introduce the numerical techniques to simulate the evolution of the systems. We present our results in Section \ref{sec:results}, and conclude in Section \ref{sec:conclusion}. 

\section{COBs and IMBHs in the Galactic center}\label{sec:numericalsetup}

\begin{figure}
  \centering
  \includegraphics[scale=1]{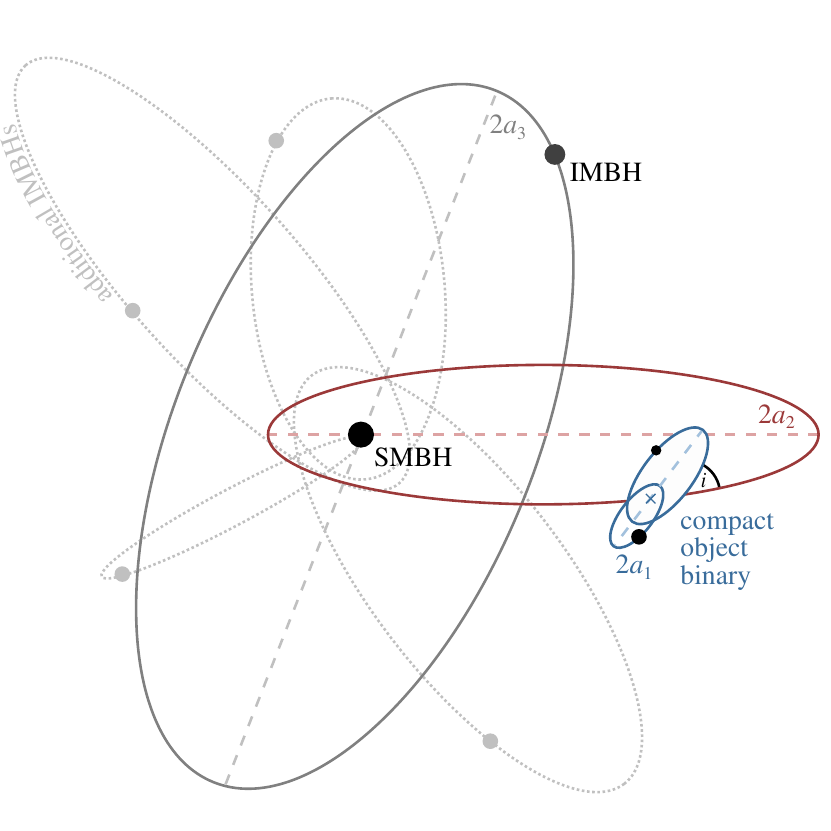}
\caption{A COB+SMBH hierarchical triple system embedded in an IMBH subsystem ($a_2 \gg a_1$). The figure is not to scale.}
\label{fig:distributions}
\end{figure}

We consider the following model shown schematically in Fig.~\ref{fig:distributions}: a tight COB (inner orbit) revolves around the central SMBH (outer orbit) constituting a hierarchical triple system. This system is perturbed by a small number of IMBHs. We use the subscripts 1 and 2 for the orbital elements of the inner and outer binary, respectively, and the subscript 3 for the orbital elements of an IMBH around the SMBH which perturbs the COB+SMBH triple system. This system represents a nested configuration of triples where the SMBH and the center of mass of the COB comprise triples with each IMBH, respectively.

We make two sets of calculations. In the first we fix the initial orbital elements of the COB+SMBH hierarchical triple and draw the orbital elements of the IMBHs in a systematic survey of simulations. In the second set, we randomly select the parameters of the COB+SMBH and the IMBHs.

\subsection{A representative system}
\label{sec:representative}

To highlight a representative system,  consider an COB+SMBH triple with  orbital elements $a_1=10 \,\mathrm{AU}$, $a_2=10^4 \,\mathrm{AU}$,
 
$\omega_1=30^{\circ}$, $\omega_2=10^{\circ}$, $\Omega_1=180^{\circ}$, $\Omega_2=0^{\circ}$, 
$e_1=0.5$, and $e_2=0.6$,
where $a$, $e$, $\omega$, and $\Omega$ denote the semi-major axis, eccentricity, argument of periapsis and ascending node, respectively (for a reference, the innermost known S-star, S2 has a semi-major axis $a_\mathrm{S2}\approx 10^3$ AU, \citealt{gillessen2017}). The masses are set to $m_\mathrm{COB}=10\,\Msun+20\,\Msun$ and $m_\mathrm{SMBH}=4.6\times10^6\mbox{ }\Msun$, consistent with the mass of the SMBH at the center of the Milky Way, e.g. \citealt{Ghez2008,Gillessen2009}.   

\subsection{Intermediate mass black holes}
\label{sec:IMBH}

For the IMBH eccentricities we assume thermal equilibrium. According to \cite{szolgyen2018}, the orbital planes of the IMBHs may be expected to settle in a disk due to vector resonant relaxation (i.e. the secular torques exerted by a cluster of objects on precessing planar orbits which drive a secular reorientation of the angular momentum direction \citealt{rauch1996,kocsis2015}) and become `anisotropically mass-segregated' (i.e. the black hole distribution becomes anisotropic because of their larger masses). However, as the expected level of anisotropy is currently poorly understood, in this work we assume isotropic initial distribution for their orientation, i.e. uniform distribution in $\omega$, $\Omega$ and $\cos i$, where $i$ is the inclination angle of the orbital planes of the IMBHs measured from the reference plane used in our simulations. We note that in some cases IMBH orbits can be so eccentric that they cross that of the outer binary (see Fig. \ref{fig:periapocenters}).    

We randomly draw five IMBHs in the relevant region of the nucleus (between $a_\mathrm{3,min}$ and $a_\mathrm{3,max}$, defined below) with mass $m_{\rm IMBH}=10^4 \,\Msun$ each. For the distribution\footnote{Note that this distribution is a function of the distance from the center ($r$), but the interval limits are meant as semi-major axes.} of the IMBHs in the galactic center we adopt the predictions of dynamical simulations by \cite{MBattisti2014}: 
\begin{equation}\label{massdistribution}
\rho_{\mathrm{IMBH}}(r)=6.2\times 10^{3}~\Msun\,\mathrm{pc^{-3}} \left(\frac{r}{1~\mathrm{pc}} \right)^{-2.32},
\end{equation}
which is consistent with the distribution produced by strong mass segregation ($\propto r^{-\alpha}$ with $2<\alpha<11/4$, see \citealt{alexanderhopman}; possibly up to $\propto r^{-3}$, \citealt{keshet2009}). The minimum separation between an IMBH and the SMBH, $a_{\mathrm{3,min}}$ is considered to be the distance at which GW-induced merger time equals the Hubble time: 
\citep{peters1964}\footnote{We evaluate this expression for $e_3=0$.}:
\begin{equation}
a_{3,\mathrm{min}}=250 \, \mathrm{AU}\times\left( \frac{m_{\mathrm{IMBH}}}{10^4 \,\Msun} \right)^{1/4} \times (1-e_3^2)^{7/8}.
\end{equation}

We specify $a_{3,\mathrm{max}}$ as the maximum distance where the secular effect of the IMBH is nonnegligible. In particular, we calculate the timescale on which IMBHs can induce significant changes in the outer orbital elements of a COB via secular (Lidov-Kozai) mechanism. The timescale of the interaction
(e.g. \citealt{naoz2016}) on which the outer binary oscillates is 
\begin{equation}\label{lktimescale}
T_{\mathrm{LK},2} \approx \frac{a_{3,\mathrm{max}}^3(1-e_3^2)^{3/2}m_{\mathrm{SMBH}}^{1/2}}{G^{1/2}a_2^{3/2}m_{\mathrm{IMBH}}},
\end{equation}
where $G$ is the gravitational constant. Note that in the nested configuration, $a_2$ represents an \textit{inner} binary and $a_{\mathrm{3}}$ is the \textit{outer} binary. We set $a_{3,\mathrm{max}}$ to be the distance at which the timescale given in Eq. (\ref{lktimescale}) equals the Hubble time. For the Galactic Center, we get
\begin{multline}
a_{3,\mathrm{max}} \approx 7.36\times10^5 \mbox{ AU } \times (1-e_3^2)^{-1/2}\times \\
\times
\left( \frac{a_2}
{\mathrm{10^4\mbox{ AU}}} \right)^{1/2} 
\left( \frac{m_{\mathrm{IMBH}}}{10^4\,\Msun} \right)^{1/3}. 
\end{multline}
For $e_{3}=1/\sqrt{2}$, the median of the thermal distribution, we get $a_{3,\mathrm{max}}=1.04\times10^6 \,\mathrm{AU}$ or $5.04\,\mathrm{pc}$.\footnote{Note that the value of $a_{\mathrm{min}}$ and $a_{\mathrm{max}}$ are calculated with $e_1=0.5$ and $e_2=0.6$, while in our simulations these values are varied.} 

In Appendix~\ref{app:crossing} we show that in a significant fraction of cases ($79\%$ for the representative system), at least one IMBH is on a radially crossing orbit with respect to the COB orbit around the SMBH. While all of the simulated systems are initially stable, they are influenced by the IMBH population secularly very efficiently as we will show.

\subsection{Allowed range of COB configurations}\label{sec:general}

Here we consider the possible COB configurations where the Lidov-Kozai effect may play a role and draw the orbital parameters randomly as follows.

Following \cite{stephan2016} and \cite{hoang2018} we choose a uniform probability distibution function for $e_1$ \citep{raghavan2010} and thermal for $e_2$ \citep{jeans1919} in the (0, 1) range. We draw $a_1$ from a loguniform distribution in the (0.1, 50) AU range, which is motivated by \cite{sana2012}. 

We draw $a_2$ from a loguniform distribution between $a_{\mathrm{2,min}}$ and $a_{\mathrm{2,max}}$. Here $a_{\mathrm{2,min}}=124\mbox{ AU}$ is chosen to be the distance where the gravitational wave inspiral time into the SMBH $T_{\mathrm{GW}}$ on which COBs are removed equals the relaxation time  $T_{\mathrm{rel}}$ on which COBs may be replenished from the outer parts of the nuclear star cluster  \citep{gondan2018}. Here \citep{peters1964}
\begin{equation}
    T_{\mathrm{GW}} \approx 0.0026 \frac{c^5a_2^4}{G^3m_{\mathrm{SMBH}}^2m_{\mathrm{COB}}},
\end{equation}
where we subsituted $e_2$ with its 
median for the thermal distribution $1/\sqrt{2}$
and \citep{Spitzer87}
\begin{equation}
    T_{\mathrm{rel}} \approx 0.34\frac{\sigma^3}{G^2nm_{*}^2\ln\Lambda},
\end{equation}
where $m_\mathrm{COB}$ and $m_\mathrm{SMBH}$ are set as in Sec.~\ref{sec:representative}, $m_{*}\approx 1~\Msun$ is the stellar mass,
$\ln\Lambda\approx 15$ is the Coulomb logarithm,
and $n\equiv\rho/m_{*}$, 
$\sigma$ is the velocity dispersion of the stellar environment \citep{kocsis2011}
\begin{equation}
\sigma \approx \left(\frac{G m_{\rm SMBH}}{a_2}\right)^{1/2} = 596 \, \mathrm{km\,s^{-1}}\left(\frac{a_2}{10^4\mbox{ AU}}\right)^{-1/2},
\end{equation}
and $\rho$ is the spatial density of stars, which is given by \citet{genzel2010} as \footnote{Note that there is a typo in Eq. (11) of \citet{kocsis2011}. The correct formula is $\sigma(r)=280\mbox{ km s}^{-1} \sqrt{0.22\mbox{ pc}/r}\sqrt{1-0.035(r/0.22\mbox{ pc})^{2.2}}$.} 
\begin{equation}
\rho \approx 8.5 \times 10^{6}~\Msun\,\mathrm{pc}^{-3} \left(\frac{ a_2}{10^4 \mbox{ AU}} \right)^{-1.3}.
\end{equation}

We set the maximum semi-major axis of the COB orbit around the SMBH arbitrarily to $a_{\mathrm{2,max}} = 2.48\times10^4$\,AU or 0.12\,pc. For this value, and for the expectation value of the inner semi-major axis $a_1=8.02\mbox{ AU}$, the binary evaporation time due to stellar encounters \citep{binneytremaine} 
\begin{equation}\label{evaporation}
    T_{\mathrm{ev}}=\frac{\sqrt{3}\sigma m_{\mathrm{COB}}}{32 \sqrt{\pi}G\rho a_1\ln \Lambda m_{*}}\approx 5.2\times 10^7\mbox{ yrs},
\end{equation}
while our maximum integration time is shorter, $500T_\mathrm{LK,1} \sim 3.3\times 10^7\mbox{ years}$, where one LK cycle lasts (analogously to Eq. (\ref{lktimescale}))
\begin{equation}\label{cycle}
T_{\mathrm{LK,1}}\approx\frac{a_2^3(1-e_2^2)^{3/2}m_{\mathrm{COB}}^{1/2}}{G^{1/2}a_1^{3/2}m_{\mathrm{SMBH}}}.
\end{equation}
Here the subscript `1' refers to the oscillations of orbit 1. Thus, binary evaporation may slightly affect our results at the upper end of the $a_2$ distribution. However, we also note that since the LK timescale increases with distance from the center more rapidly than the evaporation time ($T_\mathrm{LK,1}\propto a_2^3$, $T_\mathrm{ev}\propto a_2^{0.8}$), therefore the binaries are not disrupted before the Lidov-Kozai effect takes place in the inner regions: e.g. for $a_2=10^4$ AU the evaporation time is three orders of magnitude longer.

\section{Numerical method}\label{sec:techniques}

For investigating the dynamical effect of IMBHs on a COB+SMBH triple system, we use the code ARCHAIN \citep{archain}, which is a direct integration code based on the algorithmic regularization method \citep{algreg} and treats also post-Newtonian terms up to 2.5 order (\citealt{archainpn}; for more technical details and applications see \citealt{arcasedda2017} and \citealt{arcasedda2018}). The IMBH subsystem is evolved self-consistently, taking into account the interactions of the IMBHs with each other, the SMBH, and the COB.

The most important simplifying assumption in this work is to neglect the interactions with the surrounding nuclear star cluster, which means that our simulations lack apsidal mass precession, dynamical friction, resonant relaxation, resonant dynamical friction \citep{rauch1996} and interactions with a molecular torus. 
Many of these simplifying assumptions may fail depending on the orbital parameters of the COB, the cluster mass, torus mass, and IMBH mass: for example, 
the effects of apsidal and nodal precession may be significant \citep{chang2009,subr2009}, especially for anisotropic clusters \citep{Petrovich_Antonini2017}. 

We leave the investigation of the combined effects of the embedding stellar environment and IMBHs to future work.

In what follows, we examine the `\emph{survival probability}' in order to express how likely it is for a typical COB to remain intact around a SMBH against the perturbation of IMBHs. The survival probability is expected to decrease with time, since the more time the IMBHs perturb the COB, the more probable it is that they succeed in destroying the COB either through collision or disruption. In order to see how it depends on the orbital elements of the particular triple model (Section \ref{sec:representative}), we systematically vary the initial value of one of the orbital elements while keeping the rest fixed as follows. Masses are fixed for all simulations: $m_{\mathrm{COB}}=10\,\Msun+20\,\Msun$, $m_{\mathrm{SMBH}}=4.6\times10^6\,\Msun$, $m_\mathrm{IMBH}=10^4\,\Msun$, and so are the initial arguments of the periapsis: $\omega_{\mathrm{1}}=30^{\circ}$, $\omega_{\mathrm{2}}=10^{\circ}$. 
We run four sets of simulations as summarized in Table~\ref{tableofsbhbmergers}.
In  set [1], we vary $e_1$ from 0.0 to 0.9 while we initially fix $a_{\mathrm{1}}=10 \,\mathrm{AU}$, $a_{\mathrm{2}}=10^4 \,\mathrm{AU}$, $e_{\mathrm{2}}=0.6$ and  $i=75^{\circ}$. In set [2], $e_2$ is varied from 0.0 to 0.7 and $a_{\mathrm{1}}=10 \,\mathrm{AU}$, $a_{\mathrm{2}}=10^4 \,\mathrm{AU}$, $e_{\mathrm{1}}=0.6$,  $i=75^{\circ}$. In Set [3], we vary $i$ between $0^\circ$ and $180^\circ$, while $a_{\mathrm{1}}=10 \,\mathrm{AU}$, $a_{\mathrm{2}}=10^4 \,\mathrm{AU}$, $e_{\mathrm{1}}=0.5$ and $e_{\mathrm{2}}=0.6$. Set [4] varies $a_2/a_1$ from 600 to 900 (keeping $a_1=10$ AU fixed) and has $e_{\mathrm{1}}=0.5$, $e_{\mathrm{2}}=0.6$ and $i=75^\circ$.

\begin{table*}
\centering
\begin{tabular}{|c|c|c|c||c|c|c|c||c|c|c|c||c|c|c|c|}
\hline
\multicolumn{4}{|c||}{Set [1]} & \multicolumn{4}{c||}{Set [2]} & \multicolumn{4}{c||}{Set [3]} & \multicolumn{4}{c|}{Set [4]}\\
\hline
$e_{1}$ & runs & merg. & disr. &  $e_{2}$ & runs & merg. & disr. &  $i$ [deg] & runs & merg. & disr. &  $a_{2}$ [AU] & runs & merg. & disr. \\[1.5ex]
\hline
\hline
0.0 & 99  & 0  & 52 & 0.0 & 99  & 3 & 50  &  0   & 100 & 0 & 50 &  --  & --  & -- & -- \\ \hline
0.1 & 100 & 1  & 51 & 0.1 & 98  & 1 & 43  &  20  & 100 & 0 & 40 &  1000 & --  & -- & -- \\ \hline
0.2 & 98  & 1  & 42 & 0.2 & 100 & 6 & 42  &  40  & 98  & 0 & 46 &  2000 & --  & -- & -- \\ \hline
0.3 & 100 & 0  & 45 & 0.3 & 99  & 3 & 43  &  60  & 100 & 0 & 48 &  3000 & --  & -- & -- \\ \hline
0.4 & 100 & 3  & 46 & 0.4 & 100 & 1 & 48  &  80  & 100 & 1 & 44 &  4000 & --  & -- & -- \\ \hline
0.5 & 100 & 1  & 45 & 0.5 & 100 & 0 & 48  &  100 & 99  & 2 & 41 &  5000 & --  & -- & -- \\ \hline
0.6 & 98  & 2  & 59 & 0.6 & 100 & 0 & 41  &  120 & 100 & 0 & 47 &  6000 & 100 & 0 & 34 \\ \hline
0.7 & 98  & 3  & 46 & 0.7 & 100 & 0 & 57  &  140 & 100 & 0 & 43 &  7000 & 100 & 0 & 50 \\ \hline
0.8 & 99  & 1  & 47 & 0.8 & --  & -- & -- &  160 & 100 & 0 & 46 &  8000 & 100 & 0 & 54 \\ \hline
0.9 & 100 & 6  & 48 & 0.9 & --  & -- & -- &  180 & 100 & 0 & 48 &  9000 & 99  & 2 & 54 \\ \hline
\end{tabular}
\caption{The number of compact object binary mergers and disruptions recorded in the simulations. The fiducial COB parameters are 
$a_{1}=10 \,\mathrm{AU}$, $a_{2}=10^4 \,\mathrm{AU}$, $e_{\mathrm{2}}=0.6$, $\omega_{\mathrm{1}}=30^{\circ}$, $\omega_{\mathrm{2}}=10^{\circ}$, $i=75^{\circ}$. 
In each set, only one orbital parameter is changed as shown. For each COB, the number of all runs with different initial IMBH realizations is also indicated. Initially unstable configurations are denoted with a dash. 
\label{tableofsbhbmergers}
}
\end{table*}

For each COB orbital element choice, we run 100 simulations by randomly assigning IMBH orbital elements from the distributions given in Section \ref{sec:representative}. \footnote{We omit a small number of runs that fail due to numerical issues. For the exact number of runs see the values in Table \ref{tableofsbhbmergers}.} Each simulation is evolved for 500 Lidov-Kozai oscillations of the inner binary (Eq.~\ref{cycle}).

Note that Eq. (\ref{cycle}) gives only an order-of-magnitude estimate for the Lidov-Kozai oscillation timescale for isolated hierarchical triples. Given that our systems are perturbed, $500T_{\mathrm{LK,COB}}$ does not mean exactly 500 peaks in the $e_1$ oscillation curve.

In order to filter out systems that are initially unstable, we run a simulation for each triple parameter set without the IMBHs. We eliminate those COBs that do not survive 500 Lidov-Kozai oscillations in isolation. We note that the initial instability of the COB can also be caused by the proximity of an IMBH, therefore we also filter out those systems which are initially within the Hill sphere of any of the IMBHs. We restrict our ananlysis and conclusions to systems which are initially stable and we run a total of $\sim$ 3200 simulations. 

\section{Results}\label{sec:results}
\begin{figure}
  \centering
  \includegraphics[width=.5\textwidth]{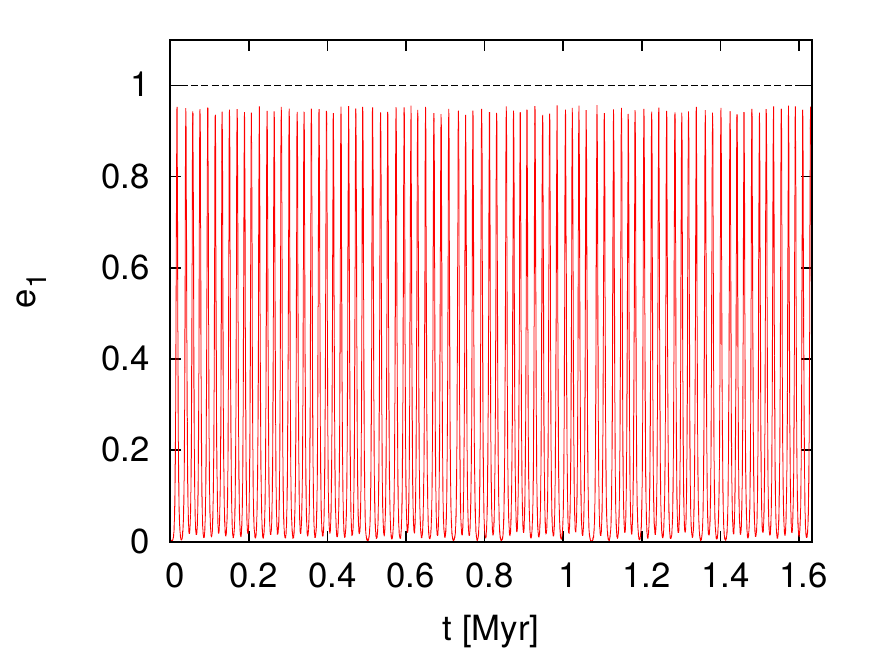}
\caption{The evolution of the internal eccentricity of a $m_{\mathrm{COB}}=10 \, \Msun+20 \, \Msun$ COB under the gravitational influence of a SMBH ($m_{\mathrm{SMBH}}=4.6\times 10^6 \,\Msun$) without any IMBHs. The initial COB parameters are given in Table~\ref{tableofsbhbmergers} (Set [1], first row).}
\label{fig:unperturbed}
\end{figure}

\begin{figure*}
\gridline{\fig{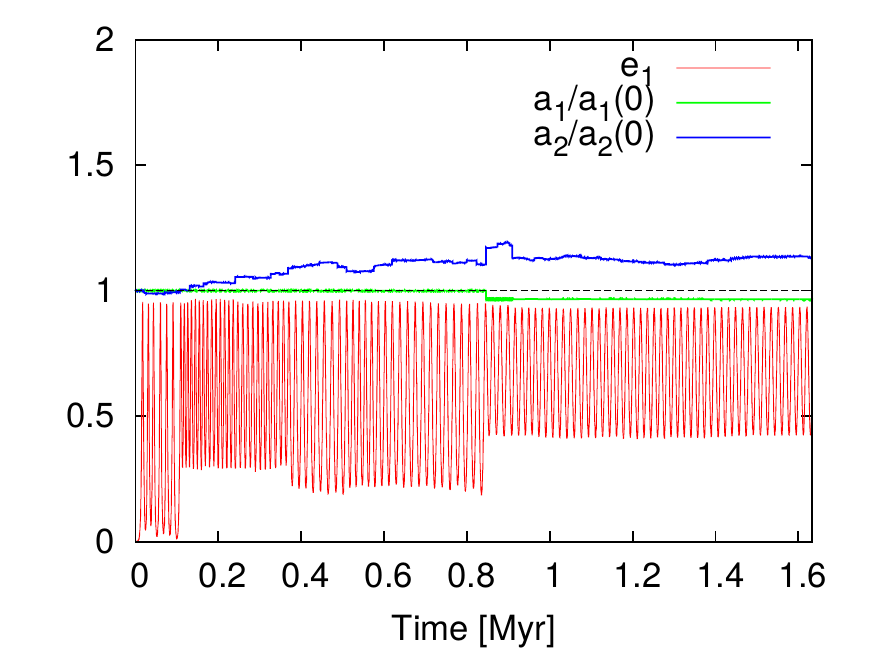}{0.5\textwidth}{(a)}
          \fig{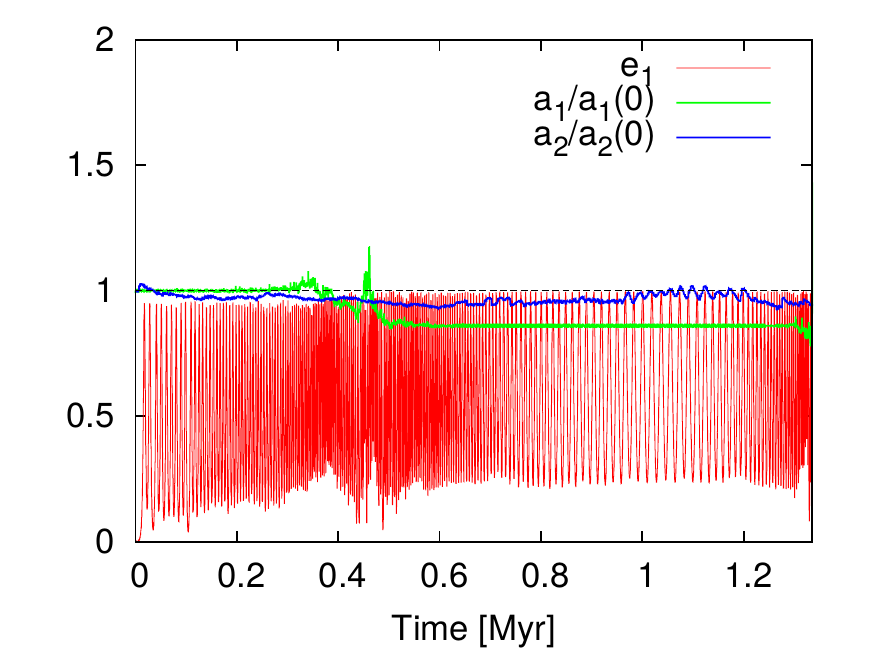}{0.5\textwidth}{(b)}}          
 \caption{
The evolution of the internal eccentricity, the inner and outer semi-major axes of a $m_{\mathrm{COB}}=10 \, \Msun+20 \, \Msun$ COB under the gravitational influence of a SMBH ($m_{\mathrm{SMBH}}=4.6\times 10^6 \,\Msun$) and five IMBHs ($m_\mathrm{IMBH}=10^4\,\Msun$ each) for two different realizations of the initial orbital parameters in the two panels. The initial COB parameters are given in Table~\ref{tableofsbhbmergers} and $e_{\mathrm{1}}=0.0$, $e_{\mathrm{2}}=0.6$ (Set [1], first row). 
IMBHs were chosen from the distribution described in Section \ref{sec:representative}. The eccentricity exhibits rapid Lidov-Kozai oscillations. In comparison, an isolated triple would produce oscillations with almost constant amplitude in the inner eccentricity and mutual inclination (see Fig. \ref{fig:unperturbed}). Deviation from  this expectation is mostly due to the presence of the IMBHs. The horizontal black dotted line represents unity. The COB in the left panel survives but that in the right panel gets disrupted as the COB eccentricity increases beyond unity.}
\label{fig:folder1}
\end{figure*}

Fig. \ref{fig:folder1} illustrates two representative examples for the eccentricity and semi-major axes evolution of a COB around a SMBH in the presence of five IMBHs. In the first case (left panel), the triple shows modulated oscillations. 
The modulation is mostly due to the quadrupole order Lidov-Kozai mechanism which produces oscillations with almost constant amplitude in the inner eccentricity and mutual inclination for isolated triples (see Fig. \ref{fig:unperturbed}).
Here the binary survives for the $500\,T_{\mathrm{LK,1}}$ integration time. In the second case (right panel), the perturbation from the IMBHs leads to the disruption of the COB within less than $500T_{\mathrm{LK,1}}$ (i.e. its inner eccentricity goes beyond unity).

\begin{figure}
  \centering
  \includegraphics[width=.5\textwidth]{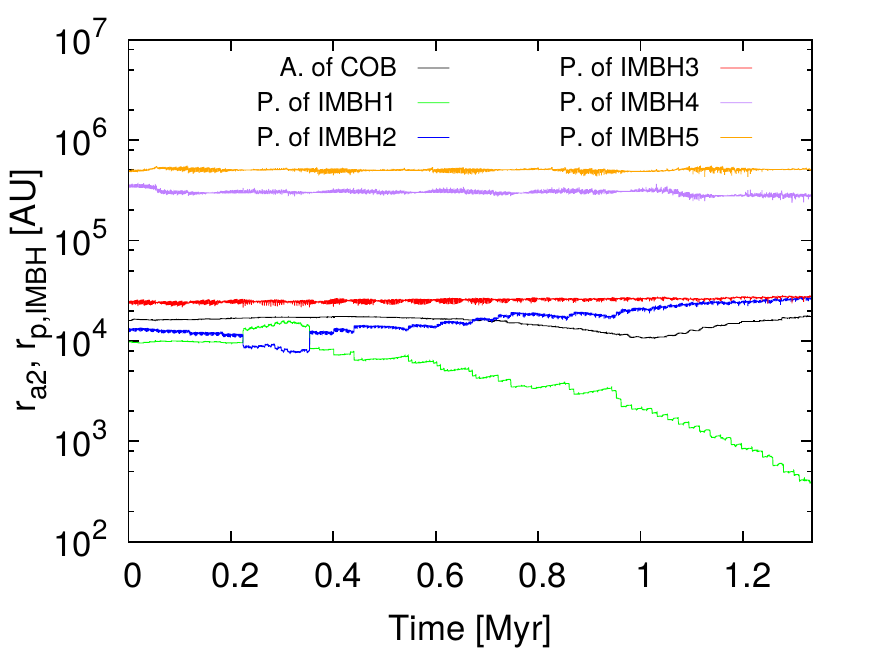}
\caption{Time evolution of the periapsides (P) of IMBHs and apoapsis (A) of the COB around the SMBH for the system shown in Fig. \ref{fig:folder1} b.}
\label{fig:periapocenters}
\end{figure}

Fig. \ref{fig:periapocenters} highlights the level of hierarchy of the COB and IMBHs orbiting the SMBH for the representative system shown in the second case of Fig. \ref{fig:folder1}. The SMBH-COB-IMBH triple is clearly not hierarchical as two of the five IMBHs are on initially radially crossing orbits with respect to the COB orbit around the SMBH. A similar non-hierarchical configuration is not uncommon. For the assumed power-law distribution for the IMBH semi-major axis (Eq. (\ref{massdistribution})) and the thermal distribution for their eccentricity, there is a $\sim$21\% probability that at least one IMBH's periapsis is smaller than the apoapsis of the outer COB orbit's.

\begin{figure*}
\gridline{\fig{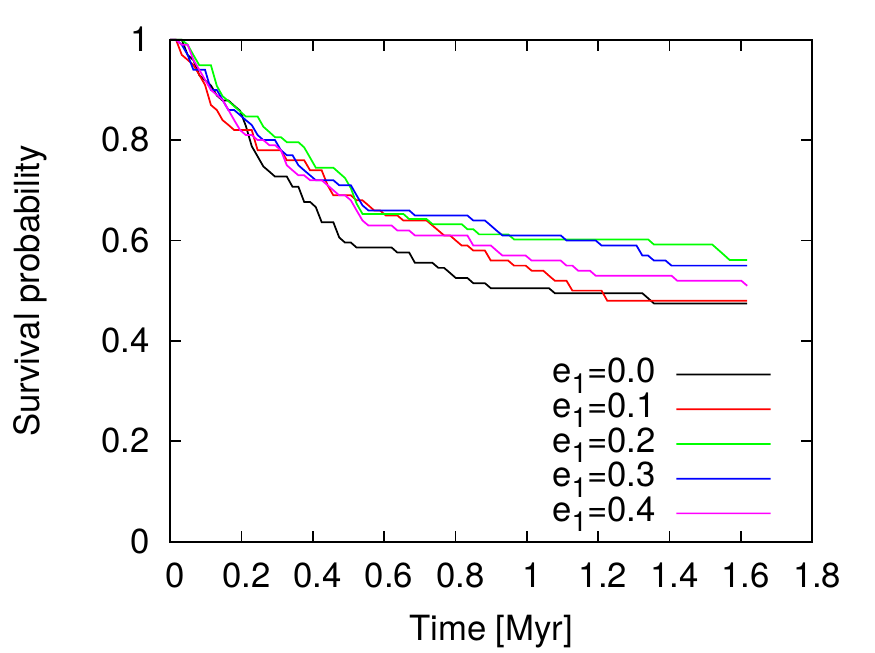}{0.5\textwidth}{(a)}
          \fig{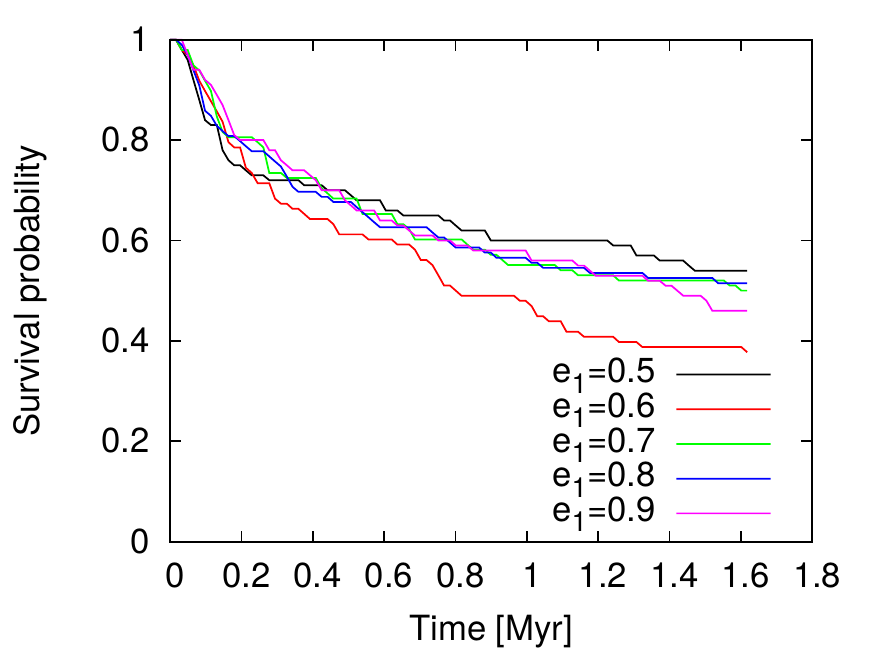}{0.5\textwidth}{(b)}} 
\caption{Survival probability as a function of time for initial inner eccentricities as shown. The other orbital parameters correspond to set [1]: $m_{\mathrm{COB}}=10+20 \, \Msun$, $m_{\mathrm{SMBH}}=4.6\times 10^6 \,\Msun$, 
$a_{1}=10 \,\mathrm{AU}$, $a_{2}=10^4 \,\mathrm{AU}$,
$e_{\mathrm{2}}=0.6$, $\omega_{\mathrm{1}}=30^{\circ}$, $\omega_{\mathrm{2}}=10^{\circ}$, $i=75^{\circ}$.
\label{fig:e1}. The two panels show the same kind of curves, it is only split in two so that the plots are not too crowded.}
\end{figure*}

\begin{figure*}
\gridline{\fig{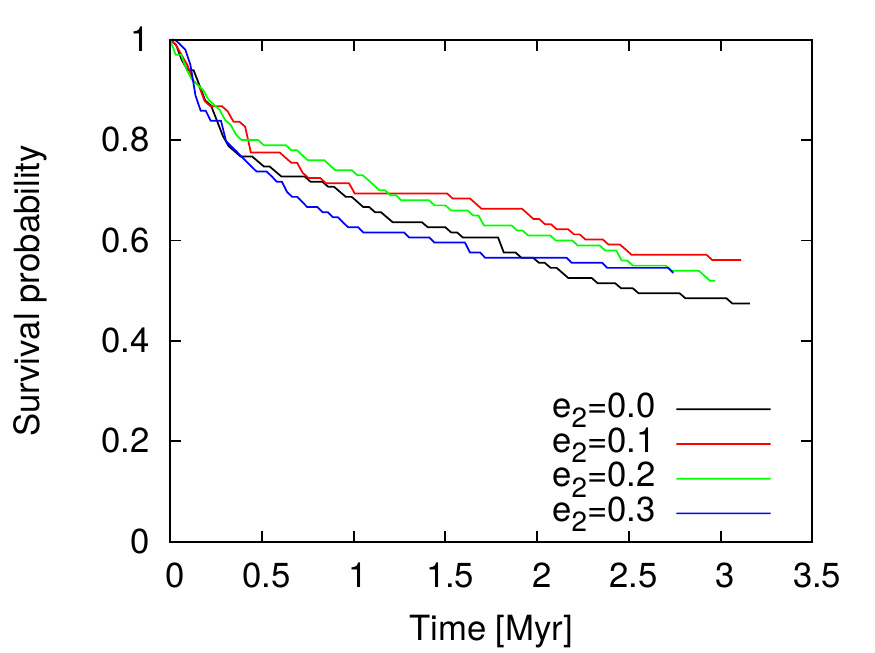}{0.5\textwidth}{(a)}
          \fig{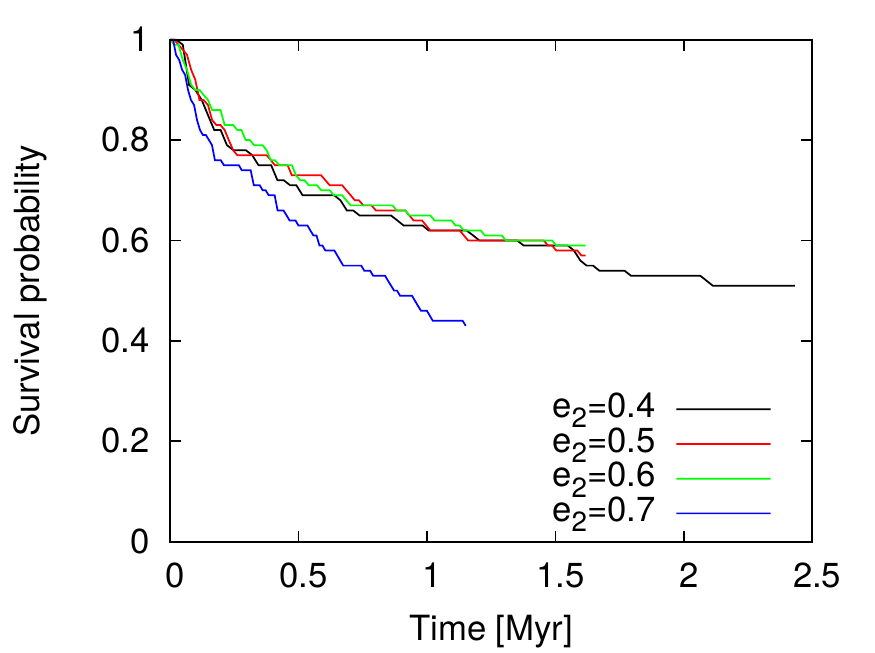}{0.5\textwidth}{(b)}} 
\caption{Same as Fig. \ref{fig:e1}, but the initial outer eccentricities are varied. The other orbital parameters correspond to set [2]: $m_{\mathrm{COB}}=10+20 \, \Msun$, $m_{\mathrm{SMBH}}=4.6\times 10^6 \,\Msun$, 
$a_{1}=10 \,\mathrm{AU}$, $a_{2}=10^4 \,\mathrm{AU}$,
$e_{\mathrm{1}}=0.6$, $\omega_{\mathrm{1}}=30^{\circ}$, $\omega_{\mathrm{2}}=10^{\circ}$, $i=75^{\circ}$}.
\label{fig:e2}
\end{figure*}

\begin{figure*}
\gridline{\fig{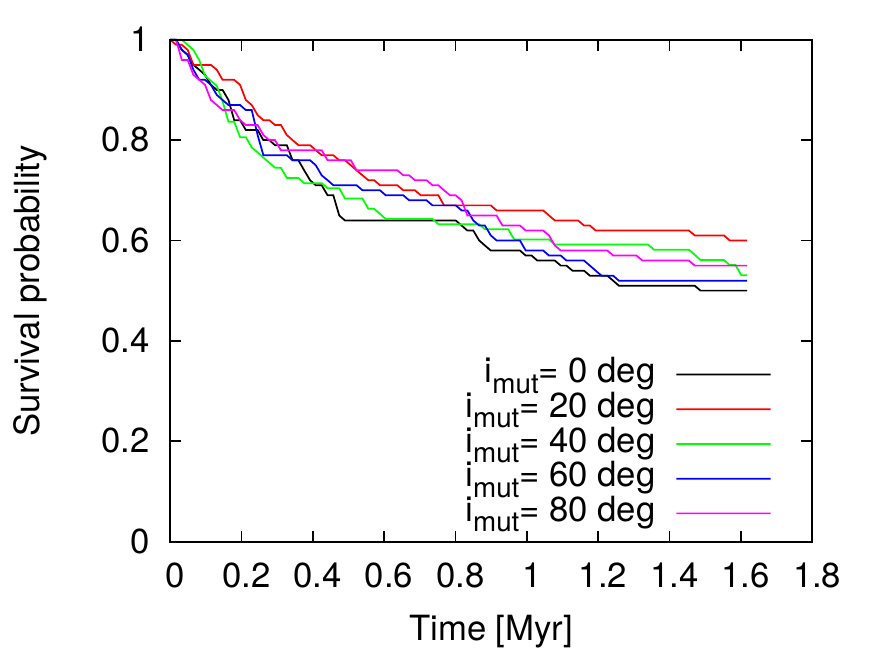}{0.5\textwidth}{(a)}
          \fig{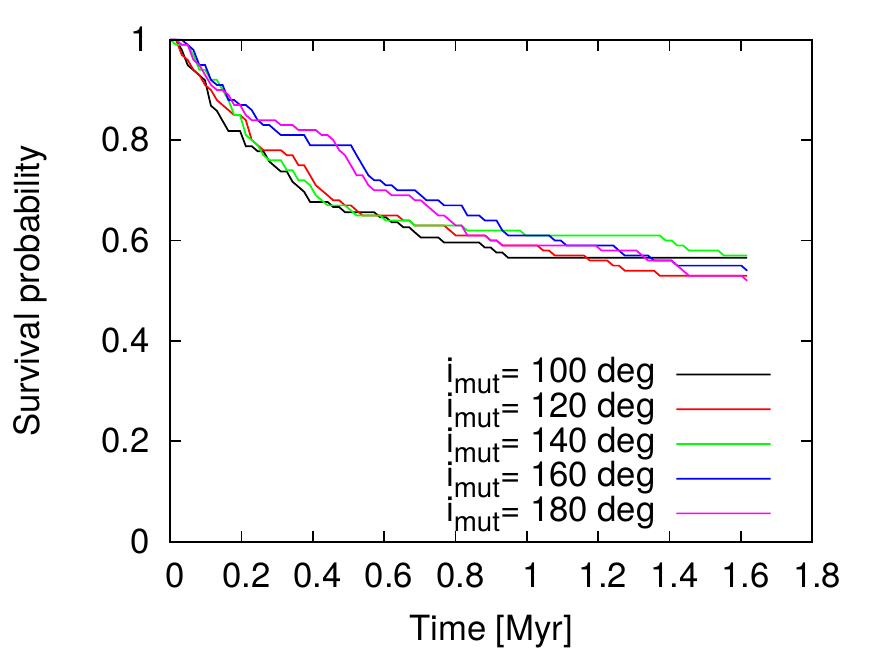}{0.5\textwidth}{(b)}} 
\caption{Same as Fig. \ref{fig:e1}, but the initial mutual inclinations are varied. The other orbital parameters correspond to set [3]: $m_{\mathrm{COB}}=10+20 \, \Msun$, $m_{\mathrm{SMBH}}=4.6\times 10^6 \,\Msun$, 
$a_{1}=10 \,\mathrm{AU}$, $a_{2}=10^4 \,\mathrm{AU}$,
$e_{\mathrm{1}}=0.5$,$e_{\mathrm{2}}=0.6$, $\omega_{\mathrm{1}}=30^{\circ}$, $\omega_{\mathrm{2}}=10^{\circ}$.}
\label{imut}
\end{figure*}

\begin{figure*}
\centering
\includegraphics[width=.5\linewidth]{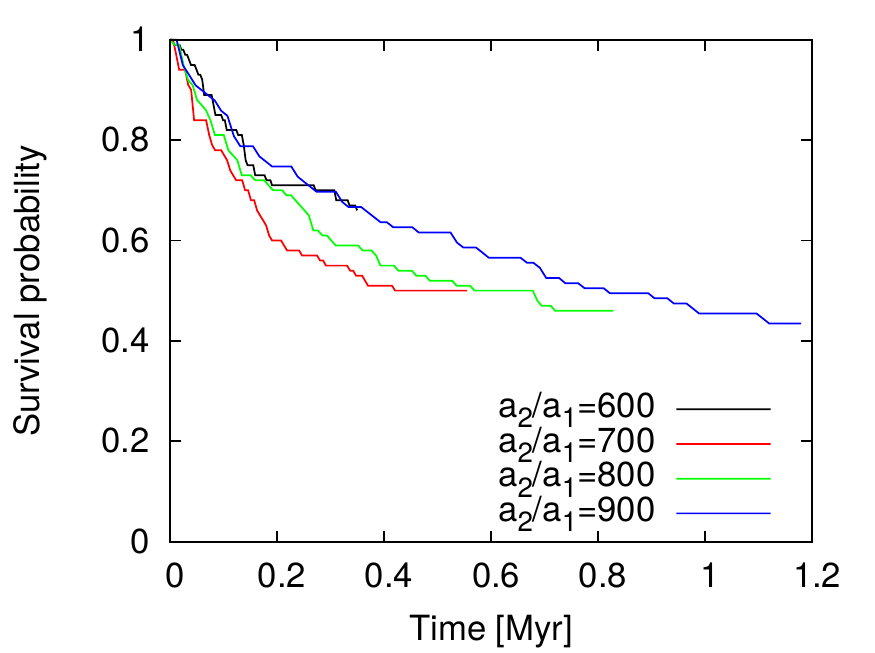}
\caption{Same as Fig. \ref{fig:e1}, but for different outer COB semi-major axis $a_2$ between $6\times 10^3\,$AU and $9\times 10^3\,$AU. The other orbital parameters correspond to simulation set [4]: $m_{\mathrm{COB}}=10+20 \, \Msun$, $m_{\mathrm{SMBH}}=4.6\times 10^6 \,\Msun$, 
$a_{1}=10 \,\mathrm{AU}$, $e_{\mathrm{1}}=0.5$, 
$e_{\mathrm{2}}=0.6$, $\omega_{\mathrm{1}}=30^{\circ}$, $\omega_{\mathrm{2}}=10^{\circ}$, $i=75^{\circ}$.}
\label{smaratio}
\end{figure*}

Figs. \ref{fig:e1}--\ref{smaratio} show the survival probability as a function of time for different orbital parameters of the triple. Apart from noise, we do not find any dependence on $e_1$, $e_2$, and $i$, as long as $e_2$ and $a_1/a_2$ are sufficiently small to avoid an immediate disruption. For $e_2 > 0.7$ or $a_2/a_1<600$  (not shown) the COB is immediately disrupted. Note that the curves terminate at different times in Figs. \ref{fig:e2} and \ref{smaratio} since the Lidov-Kozai timescale depends on these orbital elements (see Eq. (\ref{cycle})).

In Table \ref{tableofsbhbmergers} we list the number of stellar mass black hole mergers and disruptions we recorded in each of our simulation sets. The merger probability among the simulated sample shows that it is of the order of a few per cent. It is highest ($\approx$ 6\%) for $e_{1}=0.9$ or $e_{2}=0.2$ (note however the small-number statistics). The disruption probability is much larger, between $34\%$ and $60\%$.
We denote with a dash the initial parameters which lead to an initial instability even without IMBHs.

Of all COBs described in Section \ref{sec:general}, 20\% were disrupted and 2\% merged.

\subsection{The disruption mechanism}\label{discussion}

We found that the presence of 5 IMBHs within $\sim10^6$ AU ($\sim5$ pc) can significantly decrease the number of COBs (by roughly 40--50\%) within $500T_{\mathrm{LK,1}}$, which corresponds to a few $\times \,10^{5\mbox{--}6}$ years, depending on the orbital parameters (Table \ref{tableofsbhbmergers}). 
We argue that most of the COB disruptions are caused by the SMBH once the IMBHs drive the COB close to the SMBH.

\begin{figure*}
\gridline{\fig{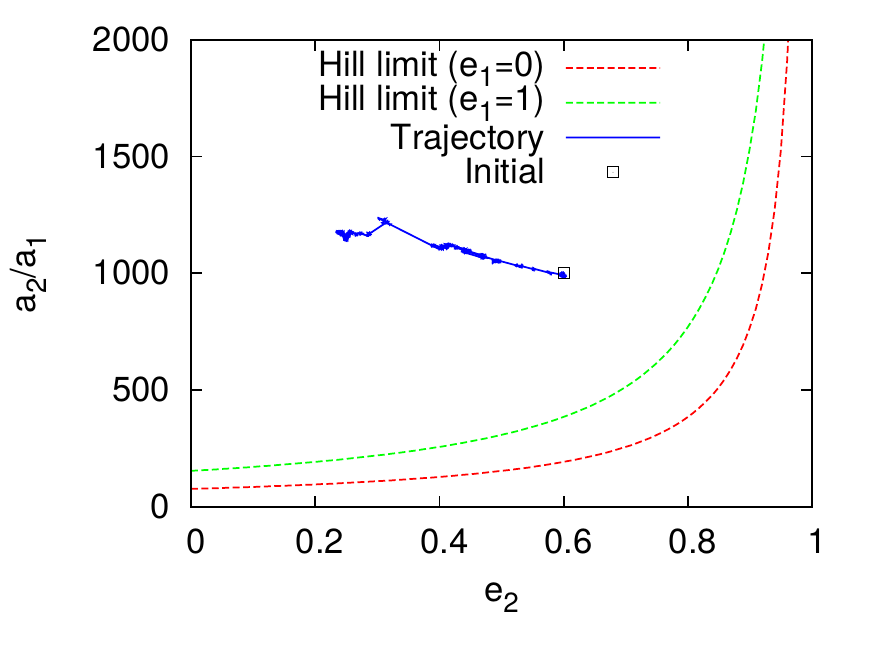}{0.5\textwidth}{(a)}
          \fig{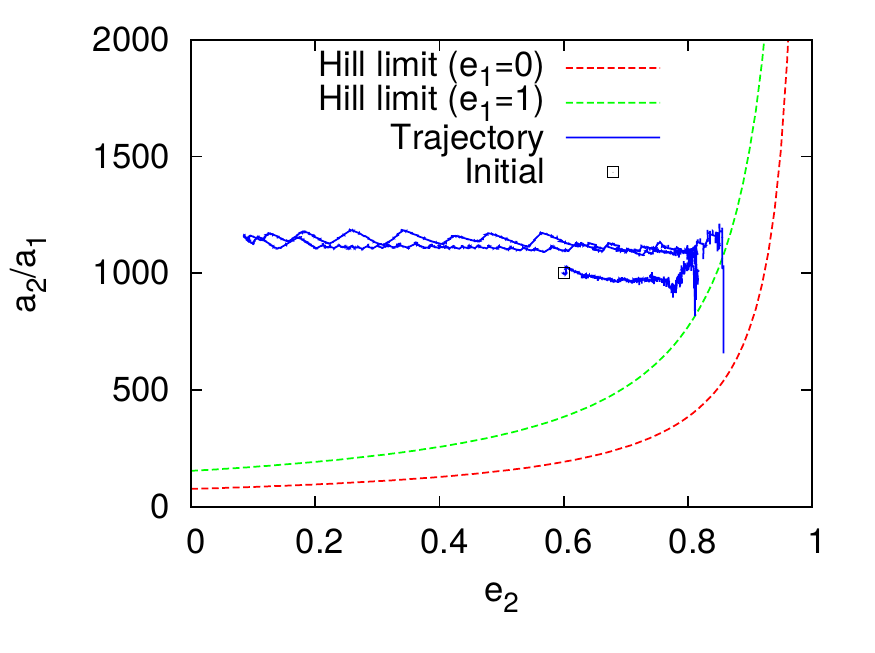}{0.5\textwidth}{(b)}} 
\caption{Trajectory of the systems in Fig. \ref{fig:folder1} in a 2D projection of the parameter space. The dotted lines represent the Hill stability limit for $e_1=0$ and $e_1=0$: the system is stable in the region above the curves and unstable below. On the left panel, the triple is perturbed deeper into the stable region. On the right panel,it is perturbed into the unstable zone ($a_2/a_1$ suddenly drops) where it eventually disrupts.}
\label{mardling}
\end{figure*}

The parameter region for Hill-unstable COBs in the vicinity of a SMBH is given by \citet{hill}
\begin{equation}\label{hill}
\frac{a_{2}}{a_{1}} < \frac{1+e_{1}}{1-e_{2}}\left( \frac{3m_{\mathrm{SMBH}}}{m_{\mathrm{COB}}} \right)^{\frac{1}{3}}
\end{equation}
\cite[see also][]{Grishin2017}.
The IMBHs may drive the triple from the stable region to the unstable one where the inner binary eventually gets disrupted by the tidal force of the SMBH.  An alternative possibility is that the COB members merge with each other. To illustrate this argument, in Fig. \ref{mardling} we plot the $a_{2}/a_1$ vs. $e_{2}$ trajectories for the systems shown in Fig. \ref{fig:folder1}. Note that Fig. \ref{mardling} shows only a 2D projection of the full parameter space, because the $e_1$-dependence of the Hill instability is weak (see the curves in the figure). The parameter space is divided into a stable and an unstable region according to Eq. (\ref{hill}). In the first case the triple system starts from a stable configuration and the IMBHs decrease $e_2$ corresponding to an even more stable configuration, and the binary remains intact within $500T_{\mathrm{LK,1}}$. However, in the second case the trajectory eventually crosses into the unstable zone where it is finally broken apart tidally by the SMBH.

The COB evolution shows that $a_2/a_1$ is mostly constant. This is expected as the effect of the IMBHs' orbit-averaged torques accumulate in a way to change the outer angular momentum of the SMBH-COB binary, i.e. $e_2$, but they cannot change the outer semi-major axis of the COB \citep{rauch1996,kocsis2015}. As in resonant relaxation, the orbit-averaged effect may be represented by smearing out the COB and the IMBH mass over their orbits. The orbital energy is conserved under the perturbation of a stationary mass distribution. However, unlike in \cite{hamers2018} where (vector) resonant relaxation slightly facilitates mergers, here it typically triggers binary disruptions.

\begin{figure*}
\gridline{\fig{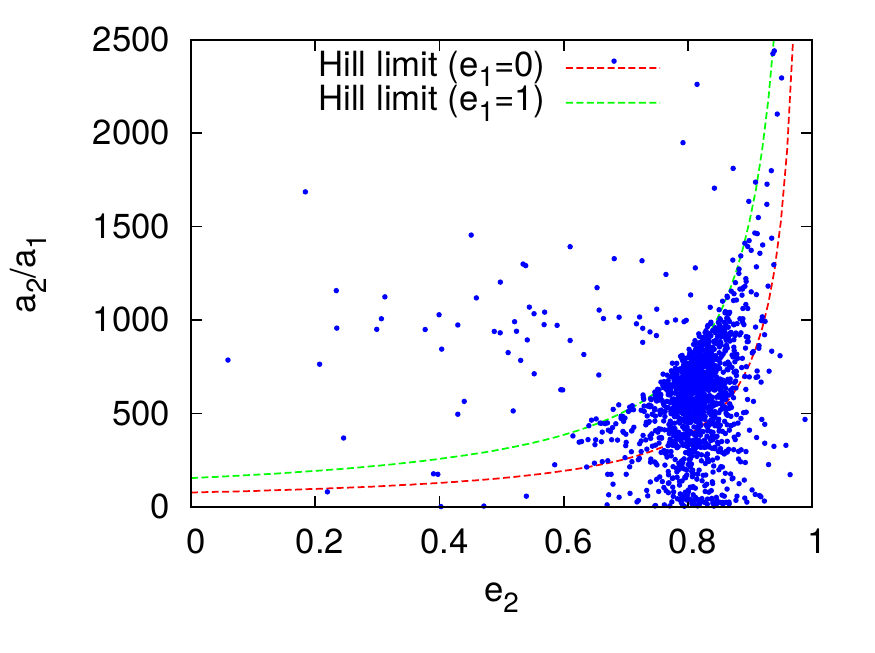}{0.5\textwidth}{(a)}
          \fig{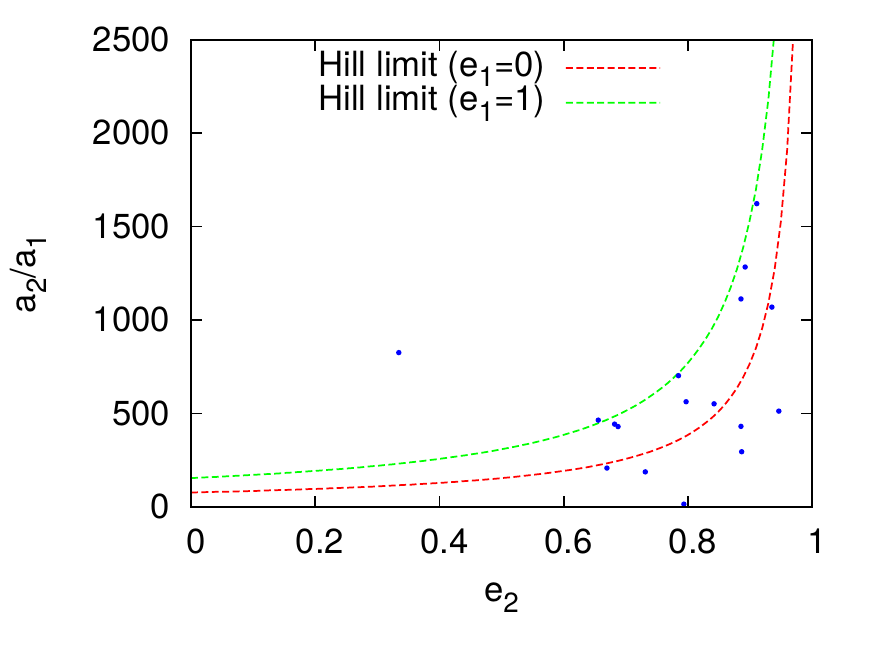}{0.5\textwidth}{(b)}}

\caption{Left: The semi-major axis ratios and outer eccentricities of the disrupted COB+SMBH triples described in Section \ref{sec:representative} shortly before their disruption for different realizations of the IMBHs in the cluster. Hill stability curve is shown for reference. Right: All of the eventually destroyed systems but for the more general COBs configurations described in Section \ref{sec:general}. In both panels, most systems lie in the Hill-unstable region, implying that the IMBHs perturb the COB in such a way that the tidal force of the SMBH finally tears it apart.
\label{fig:final}
}
\end{figure*}

In Fig. \ref{fig:final}, we plot 
the final parameter space position of the eventually destroyed systems shortly (one numerical timestep, i.e. one tenth of the orbital period) before their disruption for the representative COB system of Section~\ref{sec:representative} in the left panel and the COB distributions of Section~\ref{sec:general} in the right panel. Both panels show that most of the disrupted systems become Hill unstable. We note that the Hill disruption does not necessarily need high initial inner eccentricity ($e_1$): the eccentricity peak above unity in Fig. \ref{fig:folder1} is the consequence of being disrupted, i.e. changing the orbit from bound ellipse to an unbound hyperbola.

This implies that the IMBHs typically do not directly disrupt binaries but they play an indirect role in the COB's disruption by driving the binaries to the region where they are torn apart by the SMBH. Only in a few cases are the COBs driven into the Hill sphere of the IMBHs. 

\subsection{Hypervelocity stars}
We check whether the compact objects remain bound to the SMBH after the disruption or they escape the nuclear star clusters as hyper-velocity stars (HVSs; \citealt{Brown2015}). We found that in 99.4\% of the disrupted systems (i.e. 1526 out of 1535 simulations) both the compact objects remain bound to the SMBH. We note that this mechanism is different from that described in \cite{hills1988}: in that scenario one of the binary members is kicked out from the system and is substituted by the SMBH (exchange mechanism), while in our case both members of the binary remain bound to the SMBH. 

The fraction of escaping compact objects can be explained with the following simple argument. During the disruption of the COB, its internal energy of $E_\mathrm{COB}= G m_\mathrm{COB}/(2a_1)$ is converted to the individual orbital energies of the compact objects. If this amount of energy is larger than $E_2 = G m_\mathrm{SMBH}/(2a_2)$, i.e. $a_2/a_1 \gtrapprox m_\mathrm{SMBH}/m_\mathrm{COB}$, then at least one member of the former inner binary is ejected from the SMBH's potential well. In order to satisfy this formula and that of Hill instability (\ref{hill}) at the same time,
\begin{equation}\label{eq:HVS}
    \left(\frac{m_\mathrm{SMBH}}{m_\mathrm{COB}}\right)^{\frac{2}{3}} \lesssim \frac{1+e_1}{1-e_2} 
\end{equation}
is required for ejection. As the left-hand side is roughly $3\times10^{3}$, $e_2$ has to be very close to unity, which is satisfied only in a small part of the parameter space. The main cause of HVSs is therefore not an exchange mechanism but the rare close encounters of the IMBHs with the COBs. If we extrapolate the results for $\approx 10^6$ stars in the nucleus, of which $30\%$ are in binaries, then IMBHs may generate a few hundred HVSs in approximately 2 Myr. Thus, according to our simulations, IMBHs in the Galactic nucleus may contribute significantly to the formation of the observed HVSs. However, these estimates may be sensitive to the assumptions on the binary orbital parameters.

\subsection{Mergers}
The Lidov-Kozai mechanism is also known for its efficiency in driving the eccentricity of the inner binary to very high values at a fixed semi-major axis. This leads to the decrease of the periapsis of the inner binary, which may cause its members to eventually collide. In addition to the number of disruptions, Table~\ref{tableofsbhbmergers} lists the number of mergers as a function of the initial parameters in the four sets of simulations. Not surprisingly, the COB mergers take place for inclinations between $i=80^{\circ}$ and $100^{\circ}$, where the Lidov-Kozai effect is known to be most efficient \citep{naoz2016}.  Furthermore, mergers also favor high initial inner eccentricities ($6\%$ mergers for $e_{1}=0.9$ for the given $a_1$ and $a_2$ values, although note the low number statistics). 

For a more detailed investigation on this issue, see \citet{wang} where they put the focus on how the perturbation from two SMBHs can enhance the merger rate in a COB.  

\section{Discussion and conclusion}\label{sec:conclusion}
In this paper we have investigated the effects of IMBHs on the evolution of COBs in the nuclear star cluster around a SMBH. We found that a reservoir of 5 IMBHs may have catastrophic effects on such binaries. In many cases the IMBHs drive variations of the orbital eccentricity of the COB center of mass around the SMBH until the SMBH's tidal field disrupts the binary. The survival probability decreases by roughly 50\% within $\sim$ 500 Lidov-Kozai oscillations of the COB-SMBH system, which corresponds to less than a Myr for the S-cluster region of the Galactic Center. In most cases at least one IMBH is on a radially crossing orbit with respect to the COB's orbit around the SMBH. 

We also found that in $\lesssim1\%$ of the binary disruptions caused by the IMBH, at least one of the binary stars becomes a hypervelocity star. This may contribute significantly to the hypervelocity stars observed in the Galaxy \citep{cuihua2019}.

The perturbation of the IMBHs may also lead to the merger of the inner binary members with a few per cent probability. Chances are higher if the system is in the Lidov-Kozai inclination window (i.e. high inclinations) and if the initial inner eccentricity is also high.

Interestingly, the simulations show that the IMBHs perturb the orbits and ultimately cause their disruption very efficiently on a surprisingly short timescale (at the order of $\sim$ Myr), which is much shorter than the secular quadrupole Lidov-Kozai timescale of the SMBH-COB-IMBH systems given by
\begin{equation}\label{eq:tauLK3}
T_{\rm SMBH-COB-IMBH} \approx \frac{a_3^3 (1-e_3^2)^{3/2} m_\mathrm{SMBH}^{1/2}}{G^{1/2}a_2^{3/2}m_\mathrm{IMBH}} \approx 10^8\mbox{ yr}.
\end{equation}
The reason that the IMBHs have such a large influence on the COBs on a much shorter timescale is that in most simulations (i.e. $79\%$, see Appendix~\ref{app:crossing}) at least one IMBH is on a radially crossing orbit with respect to the COB's orbit around the SMBH. In this case the system is non-hierarchical and the secular quadrupole Lidov Kozai timescale cannot be applied. A lower limit for the IMBH's interaction timescale may be obtained by the ratio of the COB outer angular momentum and the torque exerted on it by the IMBH:
\begin{equation}
T_\mathrm{IMBH} \approx \frac{m_\mathrm{COB}\sqrt{G m_\mathrm{SMBH} a_2}}{\frac{G m_\mathrm{COB}m_\mathrm{IMBH}}{a_3^2}a_3}=\frac{a_2^{1/2}a_3 m_\mathrm{SMBH}^{1/2}}{G^{1/2}m_\mathrm{IMBH}}\approx10^5\mbox{ yr}.
\end{equation}

The timescale of the disruptions in the simulation lies between these two estimates $\tau_\mathrm{IMBH}$ and $\tau_{\rm SMBH-COB-IMBH}$.

These results are subject to the following main caveats. 
We assumed an \textit{ad hoc} number of IMBHs, namely $N=5$ IMBH of $m_{\mathrm{IMBH}}=10^4 \,\Msun$ each distributed within $\sim5$ pc of the central SMBH. While these assumptions do not violate any observations or theories about their origin, it is possible that the numbers and masses of the IMBHs are smaller. 
We also neglected the interaction with the stars of the nuclear cluster, i.e. binary evaporation, dynamical friction, and Newtonian mass precession. First, binary evaporation due to stellar encounters may decrease the binary survival rate at the upper end of the outer semi-major axis distribution (Eq.~\ref{evaporation}). Second, assuming an infinite homogeneous medium with the appropriate stellar density, the IMBH's dynamical friction timescale is estimated to be $\approx10^{5-6}$ years \citep{rasskazov2019}. However, a limited amount of stellar mass in the inner region (e.g. $1.3\times 10^4~\Msun$ at $10^4$~AU)
implies a reduced rate of dynamical friction. Indeed, \cite{MBattisti2014} find that the decay of the IMBH orbits stalls at around 0.1 pc. Third, the Newtonian mass precession timescale of the outer binary is $\approx 3\times 10^4$ years \citep{kocsis2015}. Nevertheless, the subject of mass precession is the argument of the outer pericenter. The survival probability is not very sensitive to this parameter, because it does not affect the dominant quadrupole interaction (see the so-called happy coincidence in \citet{lidov1976}), nor does it appear in the Hill instability criterion (Eq. \eqref{hill}). More generally, the assumption of neglecting the nuclear star cluster may be justified in galaxies with a massive spheroid ($M_{\rm sph}\geq 3\times 10^{10} \Msun$), where nuclear star clusters are not observed  \citep{ScottGraham2013}
and in galactic nuclei with a cored density profile \citep{2012ApJ...745...83A}.

In future work we plan to include dynamical friction on the IMBHs, Newtonian mass precession and vector resonant relaxation due to the nuclear star cluster, explore a larger region for the COBs orbit around the SMBH in the nuclear star cluster and investigate how a more or less populated IMBH reservoir would modify our conclusions. 

\section*{Acknowledgements}

We thank Manuel Arca Sedda, Giacomo Fragione and Smadar Naoz for useful discussions. This project has received funding from the European Research Council (ERC) under the European Union's Horizon 2020 research and innovation programme ERC-2014-STG under grant agreement No 638435 (GalNUC) and from the Hungarian National Research, Development, and Innovation Office grant NKFIH KH-125675. 
This research was supported in part by the National Science Foundation under Grant No. NSF PHY-1748958. 

YM acknowledges support from an NSERC grant to Ray Carlberg.
The calculations were carried out on the NIIF HPC cluster at the University of Debrecen, Hungary.

\appendix
\section{Probability of radially crossing IMBH orbits}\label{app:crossing}

Here we demonstrate that the probability of the IMBHs to be on a radially crossing orbit with respect to the COB, is in most cases high. Here radially crossing orbit refers to the case in which the periapsis of a given IMBH is smaller than the apoapsis of the COB's orbit around the SMBH. 

In this paper we adopt the results of \citet{MBattisti2014} and assume that the probability density function of the semi-major axis of the IMBHs is
\begin{equation}
\rho_a=\frac{(3+\alpha)a^{2+\alpha}}{a_\mathrm{max}^{3+\alpha}-a_\mathrm{min}^{3+\alpha}},    
\end{equation}
for $\alpha=-2.32$, and the eccentricity distribution follows
\begin{equation}
\rho_e=2e.
\end{equation}

\begin{figure*}
\gridline{\fig{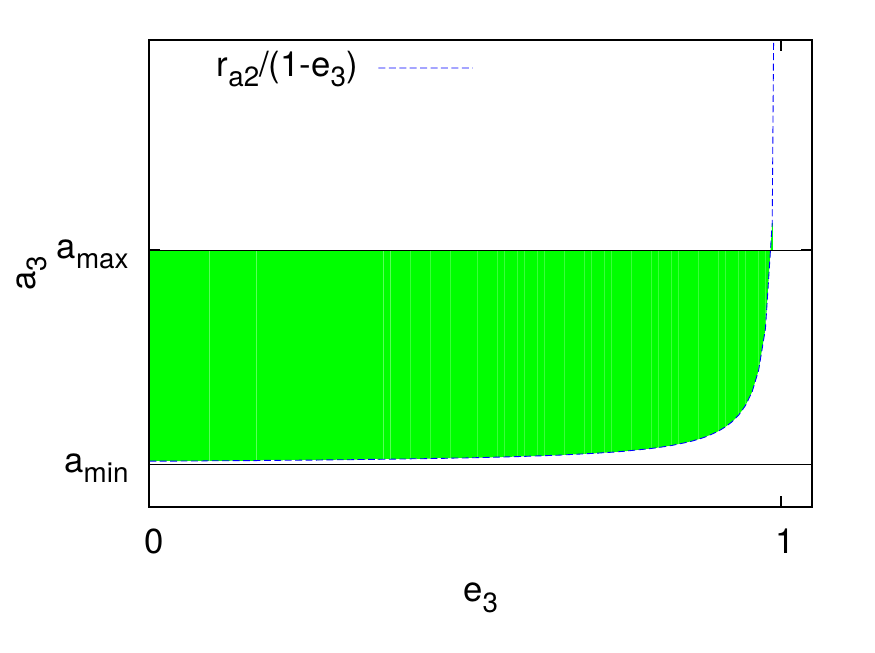}{0.5\textwidth}{(a)}
\fig{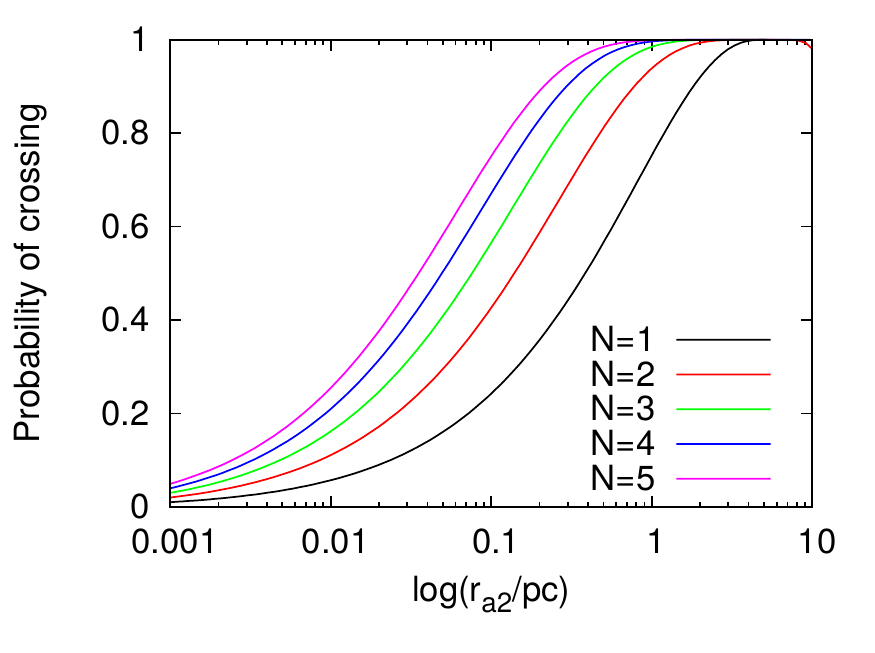}{0.5\textwidth}{(b)}} 

\caption{Left: The region of IMBH semi-major axis and eccentricity space where the IMBH orbit is radially not overlapping with the COB orbit.
Right: The probability of having at least one IMBH on a crossing orbit as a function of $\log r_{\rm a2}$ assuming a total number of $N_{\rm IMBH}$ IMBHs. The different curves refer to different IMBH number in the cluster.}
\label{appendix_fig}
\end{figure*}

Given the apoapsis of the COB's orbit around the SMBH is $r_{\rm a2} = a_2(1+e_2)$, the criterion for the IMBH to be on a crossing orbit is $a_3(1-e_3)\leq r_{\rm a2}$, or conversely, the criterion for not crossing (see the shaded area in the left panel of Fig. \ref{appendix_fig}) is $a_3(1-e_3) > r_{\rm a2}$. As the semi-major axis and the eccentricity are independent from each other, the probability of being in the $[a,a+\mathrm{d}a]$ and in the $[e,e+\mathrm{d}e]$ intervals is $\rho_a\rho_e \mathrm{d}a\mathrm{d}e$, hence the probability of not crossing is obtained by integrating $\rho_a\rho_e$ over the shaded area in the left panel of of Fig. \ref{appendix_fig}:
\begin{equation}
\overline{p}=\int_{r_{\rm a2}}^{a_\mathrm{max}} 
\rho_a
\int_0^{1-r_{\rm a2}/a}  \rho_e \mbox{}\mathrm{d}e\mbox{ }\mathrm{d}a =  \int_{r_{\rm a2}}^{a_\mathrm{max}} \frac{(3+\alpha)a^{2+\alpha}}{a_\mathrm{max}^{3+\alpha}-a_\mathrm{min}^{3+\alpha}}\left( 1-\frac{r_{\rm a2}}{a}\right)^2 \mathrm{d}a.
\end{equation}
The probability of crossing simplifies as
\begin{equation}
    p = 1-\overline{p} = 
    \frac{ 
        (r_{\rm a2}/a_{\max})^{3+\alpha} 
        }{
        1-(a_{\min}/a_{\max})^{3+\alpha}
    }
    \left[
        \frac{2}{(1+\alpha)(2+\alpha)} + \frac{6+2\alpha}{2+\alpha}
        \left(
            \frac{a_{\max}}{r_{\rm a2}}
        \right)^{2+\alpha}
        \left( 
            1 -                                 \frac{2+\alpha}{2+2\alpha}
            \frac{r_{\rm a2}}{a_{\max}}
        \right)
            -
        \left(
            \frac{a_{\min}}{r_{\rm a2}}
        \right)^{3+\alpha}
    \right]
\end{equation}

For the representative COB defined in Section~\ref{sec:representative}, we have $r_{\rm a2}=1.6\times10^4\,{\rm AU}$ and for the IMBHs we assume that $\alpha=-2.32$, $a_\mathrm{min}=0.0012120$ pc and $a_\mathrm{max}=5.04$ pc (see Section~\ref{sec:IMBH}), which yields $\overline{p}\approx0.79$, which implies that the probability of crossing for a given IMBH is $p\approx 21\%$. 

If there are $N\geq 1$ IMBHs in the star cluster, then the probability that neither one is on a crossing orbit is $\overline{p}^{N}$ and that at least one is on a crossing orbit is $1-\overline{p}^N$. For the representative COB of Section~\ref{sec:representative}, the probability of at least one IMBH out of five on a COB-crossing orbit is $\sim 70\%$. The right panel of Figure \ref{appendix_fig} shows the probability of having a radially crossing IMBH as a function of the COB apoapsis for different number of IMBHs. Given $N_{\rm IMBH}=(1,3,5)$, the COB apoaspis with respect to the SMBH must be smaller than (0.17,0.04,0.025)\,{\rm pc}, respectively, to ensure no IMBH is on a crossing orbit with at least $70\%$ probability.

\bibliography{cit}{}

\begin{thebibliography}{}
\expandafter\ifx\csname natexlab\endcsname\relax\def\natexlab#1{#1}\fi
\providecommand{\url}[1]{\href{#1}{#1}}
\providecommand{\dodoi}[1]{doi:~\href{http://doi.org/#1}{\nolinkurl{#1}}}
\providecommand{\doeprint}[1]{\href{http://ascl.net/#1}{\nolinkurl{http://ascl.net/#1}}}
\providecommand{\doarXiv}[1]{\href{https://arxiv.org/abs/#1}{\nolinkurl{https://arxiv.org/abs/#1}}}

\bibitem[{{Alexander}(2017)}]{alexander2017}
{Alexander}, T. 2017, \araa, 55, 17,
  \dodoi{10.1146/annurev-astro-091916-055306}

\bibitem[{{Alexander} \& {Hopman}(2009)}]{alexanderhopman}
{Alexander}, T., \& {Hopman}, C. 2009, \apj, 697, 1861,
  \dodoi{10.1088/0004-637X/697/2/1861}

\bibitem[{{Antonini} \& {Merritt}(2012)}]{2012ApJ...745...83A}
{Antonini}, F., \& {Merritt}, D. 2012, \apj, 745, 83,
  \dodoi{10.1088/0004-637X/745/1/83}

\bibitem[{{Antonini} \& {Perets}(2012)}]{antonini2012}
{Antonini}, F., \& {Perets}, H.~B. 2012, \apj, 757, 27,
  \dodoi{10.1088/0004-637X/757/1/27}

\bibitem[{{Arca Sedda} {et~al.}(2019){Arca Sedda}, {Askar}, \&
  {Giersz}}]{arcaseddaaskar}
{Arca Sedda}, M., {Askar}, A., \& {Giersz}, M. 2019, arXiv e-prints,
  arXiv:1905.00902.
\newblock \doarXiv{1905.00902}

\bibitem[{{Arca-Sedda} \& {Capuzzo-Dolcetta}(2017)}]{arcasedda2017}
{Arca-Sedda}, M., \& {Capuzzo-Dolcetta}, R. 2017, ArXiv e-prints.
\newblock \doarXiv{1709.05567}

\bibitem[{{Arca-Sedda} \& {Gualandris}(2018)}]{arcasedda2018}
{Arca-Sedda}, M., \& {Gualandris}, A. 2018, \mnras, 477, 4423,
  \dodoi{10.1093/mnras/sty922}

\bibitem[{{Binney} \& {Tremaine}(2008)}]{binneytremaine}
{Binney}, J., \& {Tremaine}, S. 2008, {Galactic Dynamics: Second Edition}
  (Princeton University Press)

\bibitem[{{Brown}(2015)}]{Brown2015}
{Brown}, W.~R. 2015, Annual Review of Astronomy and Astrophysics, 53, 15,
  \dodoi{10.1146/annurev-astro-082214-122230}

\bibitem[{{Chang}(2009)}]{chang2009}
{Chang}, P. 2009, \mnras, 393, 224, \dodoi{10.1111/j.1365-2966.2008.14202.x}

\bibitem[{{Du} {et~al.}(2019){Du}, {Li}, {Yan}, {Newberg}, {Shi}, {Ma}, {Chen},
  \& {Wu}}]{cuihua2019}
{Du}, C., {Li}, H., {Yan}, Y., {et~al.} 2019, \apjs, 244, 4,
  \dodoi{10.3847/1538-4365/ab328c}

\bibitem[{{Fragione} \& {Antonini}(2019)}]{Fragione_Antonini2019}
{Fragione}, G., \& {Antonini}, F. 2019, Monthly Notices of the Royal
  Astronomical Society, 488, 728, \dodoi{10.1093/mnras/stz1723}

\bibitem[{{Fragione} {et~al.}(2019{\natexlab{a}}){Fragione}, {Ginsburg}, \&
  {Loeb}}]{Fragione_Ginsburg_Loeb2019}
{Fragione}, G., {Ginsburg}, I., \& {Loeb}, A. 2019{\natexlab{a}}, arXiv
  e-prints, arXiv:1907.08008.
\newblock \doarXiv{1907.08008}

\bibitem[{{Fragione} {et~al.}(2019{\natexlab{b}}){Fragione}, {Grishin},
  {Leigh}, {Perets}, \& {Perna}}]{fragione_b}
{Fragione}, G., {Grishin}, E., {Leigh}, N. W.~C., {Perets}, H.~B., \& {Perna},
  R. 2019{\natexlab{b}}, \mnras, 488, 47, \dodoi{10.1093/mnras/stz1651}

\bibitem[{{Fragione} {et~al.}(2019{\natexlab{c}}){Fragione}, {Leigh}, \&
  {Perna}}]{fragione_a}
{Fragione}, G., {Leigh}, N. W.~C., \& {Perna}, R. 2019{\natexlab{c}}, \mnras,
  488, 2825, \dodoi{10.1093/mnras/stz1803}

\bibitem[{{Genzel} {et~al.}(2010){Genzel}, {Eisenhauer}, \&
  {Gillessen}}]{genzel2010}
{Genzel}, R., {Eisenhauer}, F., \& {Gillessen}, S. 2010, Reviews of Modern
  Physics, 82, 3121, \dodoi{10.1103/RevModPhys.82.3121}

\bibitem[{{Ghez} {et~al.}(2008){Ghez}, {Salim}, {Weinberg}, {Lu}, {Do}, {Dunn},
  {Matthews}, {Morris}, {Yelda}, {Becklin}, {Kremenek}, {Milosavljevic}, \&
  {Naiman}}]{Ghez2008}
{Ghez}, A.~M., {Salim}, S., {Weinberg}, N.~N., {et~al.} 2008, \apj, 689, 1044,
  \dodoi{10.1086/592738}

\bibitem[{{Giersz} {et~al.}(2015){Giersz}, {Leigh}, {Hypki}, {L{\"u}tzgendorf},
  \& {Askar}}]{giersz2015}
{Giersz}, M., {Leigh}, N., {Hypki}, A., {L{\"u}tzgendorf}, N., \& {Askar}, A.
  2015, \mnras, 454, 3150, \dodoi{10.1093/mnras/stv2162}

\bibitem[{{Gillessen} {et~al.}(2009){Gillessen}, {Eisenhauer}, {Trippe},
  {Alexand er}, {Genzel}, {Martins}, \& {Ott}}]{Gillessen2009}
{Gillessen}, S., {Eisenhauer}, F., {Trippe}, S., {et~al.} 2009, \apj, 692,
  1075, \dodoi{10.1088/0004-637X/692/2/1075}

\bibitem[{{Gillessen} {et~al.}(2017){Gillessen}, {Plewa}, {Eisenhauer}, {Sari},
  {Waisberg}, {Habibi}, {Pfuhl}, {George}, {Dexter}, \& {von
  Fellenberg}}]{gillessen2017}
{Gillessen}, S., {Plewa}, P.~M., {Eisenhauer}, F., {et~al.} 2017, \apj, 837,
  30, \dodoi{10.3847/1538-4357/aa5c41}

\bibitem[{{Girma} \& {Loeb}(2018)}]{girma2018}
{Girma}, E., \& {Loeb}, A. 2018, ArXiv e-prints.
\newblock \doarXiv{1807.02469}

\bibitem[{{Gond{\'a}n} {et~al.}(2018){Gond{\'a}n}, {Kocsis}, {Raffai}, \&
  {Frei}}]{gondan2018}
{Gond{\'a}n}, L., {Kocsis}, B., {Raffai}, P., \& {Frei}, Z. 2018, \apj, 860, 5,
  \dodoi{10.3847/1538-4357/aabfee}

\bibitem[{{Goodman} \& {Tan}(2004)}]{goodman2004}
{Goodman}, J., \& {Tan}, J.~C. 2004, \apj, 608, 108, \dodoi{10.1086/386360}

\bibitem[{{Grishin} {et~al.}(2017){Grishin}, {Perets}, {Zenati}, \&
  {Michaely}}]{Grishin2017}
{Grishin}, E., {Perets}, H.~B., {Zenati}, Y., \& {Michaely}, E. 2017, \mnras,
  466, 276, \dodoi{10.1093/mnras/stw3096}

\bibitem[{{Gualandris} {et~al.}(2010){Gualandris}, {Gillessen}, \&
  {Merritt}}]{gualandris2010}
{Gualandris}, A., {Gillessen}, S., \& {Merritt}, D. 2010, \mnras, 409, 1146,
  \dodoi{10.1111/j.1365-2966.2010.17373.x}

\bibitem[{{Gualandris} \& {Merritt}(2009)}]{gualandris2009}
{Gualandris}, A., \& {Merritt}, D. 2009, \apj, 705, 361,
  \dodoi{10.1088/0004-637X/705/1/361}

\bibitem[{{Hailey} {et~al.}(2018){Hailey}, {Mori}, {Bauer}, {Berkowitz},
  {Hong}, \& {Hord}}]{Hailey2018}
{Hailey}, C.~J., {Mori}, K., {Bauer}, F.~E., {et~al.} 2018, \nat, 556, 70,
  \dodoi{10.1038/nature25029}

\bibitem[{{Hamers} {et~al.}(2018){Hamers}, {Bar-Or}, {Petrovich}, \&
  {Antonini}}]{hamers2018}
{Hamers}, A.~S., {Bar-Or}, B., {Petrovich}, C., \& {Antonini}, F. 2018, ArXiv
  e-prints.
\newblock \doarXiv{1805.10313}

\bibitem[{{Hamers} \& {Samsing}(2019)}]{hamers2019}
{Hamers}, A.~S., \& {Samsing}, J. 2019, \mnras, 1557,
  \dodoi{10.1093/mnras/stz1646}

\bibitem[{{Hamilton} \& {Rafikov}(2019{\natexlab{a}})}]{hamilton1}
{Hamilton}, C., \& {Rafikov}, R.~R. 2019{\natexlab{a}}, arXiv e-prints,
  arXiv:1902.01344.
\newblock \doarXiv{1902.01344}

\bibitem[{{Hamilton} \& {Rafikov}(2019{\natexlab{b}})}]{hamilton2}
---. 2019{\natexlab{b}}, arXiv e-prints, arXiv:1902.01345.
\newblock \doarXiv{1902.01345}

\bibitem[{Hill(1878)}]{hill}
Hill, G.~W. 1878, American Journal of Mathematics, 1, 5.
\newblock \url{http://www.jstor.org/stable/2369430}

\bibitem[{{Hills}(1988)}]{hills1988}
{Hills}, J.~G. 1988, \nat, 331, 687, \dodoi{10.1038/331687a0}

\bibitem[{{Hoang} {et~al.}(2018){Hoang}, {Naoz}, {Kocsis}, {Rasio}, \&
  {Dosopoulou}}]{hoang2018}
{Hoang}, B.-M., {Naoz}, S., {Kocsis}, B., {Rasio}, F.~A., \& {Dosopoulou}, F.
  2018, \apj, 856, 140, \dodoi{10.3847/1538-4357/aaafce}

\bibitem[{{Hopman}(2009)}]{Hopman2009}
{Hopman}, C. 2009, The Astrophysical Journal, 700, 1933,
  \dodoi{10.1088/0004-637X/700/2/1933}

\bibitem[{{Jeans}(1919)}]{jeans1919}
{Jeans}, J.~H. 1919, \mnras, 79, 408, \dodoi{10.1093/mnras/79.6.408}

\bibitem[{{Keshet} {et~al.}(2009){Keshet}, {Hopman}, \&
  {Alexander}}]{keshet2009}
{Keshet}, U., {Hopman}, C., \& {Alexander}, T. 2009, \apjl, 698, L64,
  \dodoi{10.1088/0004-637X/698/1/L64}

\bibitem[{{Kocsis} {et~al.}(2012){Kocsis}, {Ray}, \& {Portegies
  Zwart}}]{kocsis2012}
{Kocsis}, B., {Ray}, A., \& {Portegies Zwart}, S. 2012, \apj, 752, 67,
  \dodoi{10.1088/0004-637X/752/1/67}

\bibitem[{{Kocsis} \& {Tremaine}(2011)}]{kocsis2011}
{Kocsis}, B., \& {Tremaine}, S. 2011, \mnras, 412, 187,
  \dodoi{10.1111/j.1365-2966.2010.17897.x}

\bibitem[{{Kocsis} \& {Tremaine}(2015)}]{kocsis2015}
---. 2015, \mnras, 448, 3265, \dodoi{10.1093/mnras/stv057}

\bibitem[{{Leigh} {et~al.}(2014){Leigh}, {L{\"u}tzgendorf}, {Geller},
  {Maccarone}, {Heinke}, \& {Sesana}}]{leigh}
{Leigh}, N. W.~C., {L{\"u}tzgendorf}, N., {Geller}, A.~M., {et~al.} 2014,
  \mnras, 444, 29, \dodoi{10.1093/mnras/stu1437}

\bibitem[{{Lidov} \& {Ziglin}(1976)}]{lidov1976}
{Lidov}, M.~L., \& {Ziglin}, S.~L. 1976, Celestial Mechanics, 13, 471,
  \dodoi{10.1007/BF01229100}

\bibitem[{{Lithwick} \& {Naoz}(2011)}]{lithwick}
{Lithwick}, Y., \& {Naoz}, S. 2011, \apj, 742, 94,
  \dodoi{10.1088/0004-637X/742/2/94}

\bibitem[{{Madau} \& {Rees}(2001)}]{madau2001}
{Madau}, P., \& {Rees}, M.~J. 2001, \apjl, 551, L27, \dodoi{10.1086/319848}

\bibitem[{{Mastrobuono-Battisti} {et~al.}(2014){Mastrobuono-Battisti},
  {Perets}, \& {Loeb}}]{MBattisti2014}
{Mastrobuono-Battisti}, A., {Perets}, H.~B., \& {Loeb}, A. 2014, \apj, 796, 40,
  \dodoi{10.1088/0004-637X/796/1/40}

\bibitem[{{McKernan} {et~al.}(2014){McKernan}, {Ford}, {Kocsis}, {Lyra}, \&
  {Winter}}]{McKernan2014}
{McKernan}, B., {Ford}, K.~E.~S., {Kocsis}, B., {Lyra}, W., \& {Winter}, L.~M.
  2014, \mnras, 441, 900, \dodoi{10.1093/mnras/stu553}

\bibitem[{{McKernan} {et~al.}(2012){McKernan}, {Ford}, {Lyra}, \&
  {Perets}}]{mckernan2012}
{McKernan}, B., {Ford}, K.~E.~S., {Lyra}, W., \& {Perets}, H.~B. 2012, \mnras,
  425, 460, \dodoi{10.1111/j.1365-2966.2012.21486.x}

\bibitem[{{Merritt} {et~al.}(2009){Merritt}, {Gualandris}, \&
  {Mikkola}}]{merritt2009}
{Merritt}, D., {Gualandris}, A., \& {Mikkola}, S. 2009, The Astrophysical
  Journal Letters, 693, L35, \dodoi{10.1088/0004-637X/693/1/L35}

\bibitem[{{Mezcua}(2017)}]{mezcua2017}
{Mezcua}, M. 2017, International Journal of Modern Physics D, 26, 1730021,
  \dodoi{10.1142/S021827181730021X}

\bibitem[{{Mikkola} \& {Aarseth}(1990)}]{algreg}
{Mikkola}, S., \& {Aarseth}, S.~J. 1990, Celestial Mechanics and Dynamical
  Astronomy, 47, 375

\bibitem[{{Mikkola} \& {Merritt}(2008)}]{archainpn}
{Mikkola}, S., \& {Merritt}, D. 2008, \aj, 135, 2398,
  \dodoi{10.1088/0004-6256/135/6/2398}

\bibitem[{{Mikkola} \& {Tanikawa}(1999)}]{archain}
{Mikkola}, S., \& {Tanikawa}, K. 1999, \mnras, 310, 745,
  \dodoi{10.1046/j.1365-8711.1999.02982.x}

\bibitem[{{Naoz}(2016)}]{naoz2016}
{Naoz}, S. 2016, Annual Review of Astronomy and Astrophysics, 54, 441,
  \dodoi{10.1146/annurev-astro-081915-023315}

\bibitem[{{Neumayer} {et~al.}(2020){Neumayer}, {Seth}, \&
  {Boeker}}]{neumayer2020}
{Neumayer}, N., {Seth}, A., \& {Boeker}, T. 2020, arXiv e-prints,
  arXiv:2001.03626.
\newblock \doarXiv{2001.03626}

\bibitem[{Peters(1964)}]{peters1964}
Peters, P.~C. 1964, Phys. Rev., 136, B1224, \dodoi{10.1103/PhysRev.136.B1224}

\bibitem[{{Petrovich} \& {Antonini}(2017{\natexlab{a}})}]{petrovich2017}
{Petrovich}, C., \& {Antonini}, F. 2017{\natexlab{a}}, \apj, 846, 146,
  \dodoi{10.3847/1538-4357/aa8628}

\bibitem[{{Petrovich} \&
  {Antonini}(2017{\natexlab{b}})}]{Petrovich_Antonini2017}
---. 2017{\natexlab{b}}, The Astrophysical Journal, 846, 146,
  \dodoi{10.3847/1538-4357/aa8628}

\bibitem[{{Pfuhl} {et~al.}(2014){Pfuhl}, {Alexander}, {Gillessen}, {Martins},
  {Genzel}, {Eisenhauer}, {Fritz}, \& {Ott}}]{pfuhl2014}
{Pfuhl}, O., {Alexander}, T., {Gillessen}, S., {et~al.} 2014, \apj, 782, 101,
  \dodoi{10.1088/0004-637X/782/2/101}

\bibitem[{{Portegies Zwart} {et~al.}(2006){Portegies Zwart}, {Baumgardt},
  {McMillan}, {Makino}, {Hut}, \& {Ebisuzaki}}]{pzwart2006}
{Portegies Zwart}, S.~F., {Baumgardt}, H., {McMillan}, S.~L.~W., {et~al.} 2006,
  \apj, 641, 319, \dodoi{10.1086/500361}

\bibitem[{{Raghavan} {et~al.}(2010){Raghavan}, {McAlister}, {Henry}, {Latham},
  {Marcy}, {Mason}, {Gies}, {White}, \& {ten Brummelaar}}]{raghavan2010}
{Raghavan}, D., {McAlister}, H.~A., {Henry}, T.~J., {et~al.} 2010, \apjs, 190,
  1, \dodoi{10.1088/0067-0049/190/1/1}

\bibitem[{{Rasskazov} \& {Kocsis}(2019)}]{rasskazov2019}
{Rasskazov}, A., \& {Kocsis}, B. 2019, arXiv e-prints.
\newblock \doarXiv{1902.03242}

\bibitem[{{Rauch} \& {Tremaine}(1996)}]{rauch1996}
{Rauch}, K.~P., \& {Tremaine}, S. 1996, Journal of the Royal Astronomical
  Society of Canada, 90, 334

\bibitem[{{Sana} {et~al.}(2012){Sana}, {de Mink}, {de Koter}, {Langer},
  {Evans}, {Gieles}, {Gosset}, {Izzard}, {Le Bouquin}, \&
  {Schneider}}]{sana2012}
{Sana}, H., {de Mink}, S.~E., {de Koter}, A., {et~al.} 2012, Science, 337, 444,
  \dodoi{10.1126/science.1223344}

\bibitem[{{Scott} \& {Graham}(2013)}]{ScottGraham2013}
{Scott}, N., \& {Graham}, A.~W. 2013, \apj, 763, 76,
  \dodoi{10.1088/0004-637X/763/2/76}

\bibitem[{{Spitzer}(1987)}]{Spitzer87}
{Spitzer}, L. 1987, {Dynamical evolution of globular clusters}

\bibitem[{{Stephan} {et~al.}(2016){Stephan}, {Naoz}, {Ghez}, {Witzel},
  {Sitarski}, {Do}, \& {Kocsis}}]{stephan2016}
{Stephan}, A.~P., {Naoz}, S., {Ghez}, A.~M., {et~al.} 2016, \mnras, 460, 3494,
  \dodoi{10.1093/mnras/stw1220}

\bibitem[{{Stephan} {et~al.}(2017){Stephan}, {Naoz}, \&
  {Zuckerman}}]{stephan2017}
{Stephan}, A.~P., {Naoz}, S., \& {Zuckerman}, B. 2017, \apjl, 844, L16,
  \dodoi{10.3847/2041-8213/aa7cf3}

\bibitem[{{Sz{\"o}lgy{\'e}n} \& {Kocsis}(2018)}]{szolgyen2018}
{Sz{\"o}lgy{\'e}n}, {\'A}., \& {Kocsis}, B. 2018, Physical Review Letters, 121,
  101101, \dodoi{10.1103/PhysRevLett.121.101101}

\bibitem[{{The LIGO Scientific Collaboration} {et~al.}(2018){The LIGO
  Scientific Collaboration}, {the Virgo Collaboration}, {Abbott}, {Abbott},
  {Abbott}, {Abraham}, {Acernese}, {Ackley}, {Adams}, \& {Adhikari}}]{o2}
{The LIGO Scientific Collaboration}, {the Virgo Collaboration}, {Abbott},
  B.~P., {et~al.} 2018, arXiv e-prints, arXiv:1811.12907.
\newblock \doarXiv{1811.12907}

\bibitem[{{Trani} {et~al.}(2019){Trani}, {Fujii}, \&
  {Spera}}]{Trani_Fujii_Spera2019}
{Trani}, A.~A., {Fujii}, M.~S., \& {Spera}, M. 2019, The Astrophysical Journal,
  875, 42, \dodoi{10.3847/1538-4357/ab0e70}

\bibitem[{{{\v S}ubr} {et~al.}(2009){{\v S}ubr}, {Schovancov{\'a}}, \&
  {Kroupa}}]{subr2009}
{{\v S}ubr}, L.~., {Schovancov{\'a}}, J., \& {Kroupa}, P. 2009, \aap, 496, 695,
  \dodoi{10.1051/0004-6361:200811075}

\bibitem[{{Wang} {et~al.}(2019){Wang}, {Leigh}, {Sesana}, \& {Perna}}]{wang}
{Wang}, Y.-H., {Leigh}, N., {Sesana}, A., \& {Perna}, R. 2019, \mnras, 482,
  3206, \dodoi{10.1093/mnras/sty2866}

\bibitem[{{Yu} \& {Tremaine}(2003)}]{yu2003}
{Yu}, Q., \& {Tremaine}, S. 2003, \apj, 599, 1129, \dodoi{10.1086/379546}

\end{thebibliography}
\bibliographystyle{aasjournal}

\end{document}